\documentclass[a4paper,fleqn,usenatbib]{mnras}
\usepackage{CJK}
\usepackage{newtxtext,newtxmath}
\usepackage[T1]{fontenc}
\usepackage{ae,aecompl}
\usepackage[]{units}
\usepackage{graphicx}
\usepackage{amssymb}
\usepackage{amsmath}
\usepackage{threeparttable}
\usepackage{hyperref}
\usepackage{color}
\usepackage{hhline}
\usepackage{subfigure}
\usepackage{lineno}

\mathchardef\mhyphen="2D

\title[Covariance Matrix]{Decorrelating the errors of the galaxy correlation function with compact transformation matrices}

\author[S. Yuan and D. J. Eisenstein]{
Sihan Yuan,$^{1}$\thanks{E-mail: sihan.yuan@cfa.harvard.edu}
and Daniel J. Eisenstein$^{1}$
\\
$^{1}$Harvard-Smithsonian Center for Astrophysics, 60 Garden St., Cambridge, MA 02138, USA
}

\date{Accepted XXX. Received YYY; in original form ZZZ}

\pubyear{2018}

\begin{document}
\label{firstpage}
\pagerange{\pageref{firstpage}--\pageref{lastpage}}
\maketitle

\begin{abstract} 
Covariance matrix estimation is a persistent challenge for cosmology, often requiring a large number of synthetic mock catalogues.
The off-diagonal components of the covariance matrix also make it difficult to show representative error bars on the 2-point correlation function (2PCF), since errors computed from the diagonal values of the covariance matrix greatly underestimate the uncertainties. 
We develop a routine for decorrelating the projected and anisotropic 2PCF with simple and scale-compact transformations on the 2PCF. These transformation matrices are modeled after the Cholesky decomposition and the symmetric square root of the Fisher matrix. Using mock catalogues, we show that the transformed projected and anisotropic 2PCF recover the same structure as the original 2PCF, while producing largely decorrelated error bars. Specifically, we propose simple Cholesky based transformation matrices that suppress the off-diagonal covariances on the projected 2PCF by ${\sim} 95\%$ and that on the anisotropic 2PCF by ${\sim} 87\%$. These transformations also serve as highly regularized models of the Fisher matrix, compressing the degrees of freedom so that one can fit for the Fisher matrix with a much smaller number of mocks. 
\end{abstract}
\begin{keywords}
cosmology: large-scale structure of Universe -- cosmology: dark matter -- galaxies: haloes -- methods: analytical 
\end{keywords}

\section{Introduction}
\label{sec:intro}

The 2-point correlation function (2PCF) is one of the most powerful cosmological probes. It quantifies the excess probability of finding one galaxy within a specified distance of another galaxy relative to a random distribution of galaxies. For the case of a Gaussian random field, the 3-point correlation function and higher order connected correlation functions are zero, and the 2PCF encapsulates the full statistical properties of the field, and thus contains all information on the cosmological parameters \citep[e.g.][]{1980Peebles, 2013Wang, 2017Alam}. Even on the small scale, where the Gaussian random field assumption no longer holds, the 2PCF still serves as an essential probe of the galaxy formation models and galaxy-halo connection models \citep[e.g.][]{2004Tegmark, 2009Zheng, 2017Zhai}. 

However, extracting parameter constraints from observed 2PCFs requires accurate determination of the covariance matrix for use in the likelihood function. Traditionally, the covariance matrix is determined with a large number of reasonably-accurate mocks. The consequences of having an insufficient number of mocks are well documented in the literature \citep{2013Dodelson, 2013Taylor,2014Percival, 2018OConnell}.
Generally speaking, when $n_{\rm mock}$ mocks are used to generate the covariance matrix for a correlation function in $n_{\rm bin}$ bins, the noise in the covariance matrix scales as $n_{\rm bin}/n_{\rm mock}$, corresponding to a fractional increase in the uncertainty in the cosmological parameters, relative to an ideal measurement, of $\mathcal{O}(1/(n_{\rm{mock}}- n_{\rm{bin}}))$. Noise in the covariance matrix also leads to a biased estimate of the inverse covariance matrix. 
We propose a set of simple linear transformations to the 2PCF that are compact in real space and largely decorrelates the different separation bins. These transformations provide a template model for the inverse covariance matrix with only a few parameters. Using this model, one can use a much smaller number of mocks to fit the inverse covariance matrix and avoid the inversion-induced bias in the fit. Then one can run mocks using these constraints to generate meaningful error bars.


The heavy correlation between different separation bins in the 2PCF means that plots of the 2PCF can be difficult to interpret. For example, amplitude fluctuations in poorly constrained Fourier modes of very low wavenumbers cause the entire 2PCF to shift up and down. Traditionally, the error bars plotted on the 2PCF are computed from only the diagonal components of the covariance matrix, thus underestimating the uncertainties in the 2PCF. Our linear transformations to the 2PCF provide a new basis in which the transformed 2PCF has a close-to-diagonal covariance matrix, 
Then one can compute error bars from the new diagonal components that do capture most of the uncertainties in the transformed 2PCF. 
Before we proceed, we note that this problem is a purely pedagogical one as it would not be a problem while fitting cosmological parameters since one would always use the full covariance matrix. 


Such transformations are extensively discussed in \citet{2000aHamilton, 2000bHamilton}.  There are an infinite number of choices of bases that will produce diagonal covariance matrices, but the challenge is to find a transformation that is simple and compact in scale. It needs to be simple in the sense that it has few parameters so that it does not require a large number of mocks to determine. It needs to be compact in real space such that it induces minimal scale mixing since it is hard to interpret a transformed 2PCF that mixes a wide range of scales. In section~6.5 of \citet{2014Anderson}, a simply-defined and compact transformation is proposed for the monopole 2PCF at large scales. The paper shows dramatic suppression to the off-diagonal terms of the covariance matrix and a largely decorrelated formulation of the 2PCF at large scales.

In this paper, we present simple and compact transformations for both the projected 2PCF and the anisotropic 2PCF that decorrelate them on the small scale. We model the transformation matrix both on the symmetric square root of the Fisher matrix, as done in \citet{2014Anderson}, and on the Cholesky decomposition of the Fisher matrix. We note that the values in the covariance matrix and the transformation matrix are specific to the galaxy sample and the binning used. The procedure presented in this paper should be regarded as a guideline and not as a universal formula. We also clarify that, we use the name ``Fisher matrix" to refer to the inverse covariance matrix throughout this paper.

This paper is structured as the following. In Section~\ref{sec:theory}, we present the special case where $\xi(r)\propto r^{-2}$. In this case, the exact form of the covariance matrix can be computed analytically, and the inverse is strictly tridiagonal, enabling particularly compact transformation matrices. In Section~\ref{sec:simulation}, we present the N-body simulations and the galaxy-halo connection models that we use for generating the mock galaxy catalogs and the correlation functions. In Section~\ref{sec:projected} and Section~\ref{sec:anisotropic}, we present the methodology for decorrelating the projected 2PCF and the anisotropic 2PCF, respectively. Finally, in Section~\ref{sec:conclusions}, we present some discussion and conclusive remarks. 

\section{A theoretical pre-text}
\label{sec:theory}

\newcommand{\vecr}{\vec{r}}
\newcommand{\vk}{\vec{k}}

We start by presenting a curious result in a common approximation for large-scale
structure that the covariance matrix of the spherically-averaged
correlation function has a tridiagonal inverse if $\xi(r)\propto
r^{-2}$ and we neglect boundary effects, shot noise, and any
non-Gaussianity of the field.  This result can be proven from a
hidden application of Gauss's law.  A tridiagonal Fisher matrix
implies a Cholesky decomposition that has only one sub-diagonal and
hence that the quantity $d\xi/dr$ has a diagonal covariance in this
limit.  While true large-scale structure and real surveys do not 
obey the exact assumptions of this, it is possible
that this result might offer some opportunities in pre-whitening data
sets or in the interpretation of some previously observed results 
about the near-tridiagonal structure of the Fisher matrix in surveys.

We consider the correlation function averaged into spherical shells,
labeled as $a=1\ldots N$.  We assume periodic boundary conditions,
so as to avoid any boundaries.  We assume a Gaussian random field
and neglect shot noise.  We neglect redshift distortions, so the
statistical correlations are isotropic.  We then wish to compute
the covariance matrix of these shells of the correlation function.

If we start by computing the correlation function at specific 3-d
vector separations, $\vecr_a$, then we have 
\begin{equation}
\xi(\vecr_a) = \int \frac{d^3k}{(2\pi)^3} P(\vk) e^{i\vk\cdot \vecr_a}
\end{equation}
where $P(\vk)$ is the power spectrum. It is then easy to prove that the covariance between the correlation
function at two separate vectors is
\begin{align}
C_{ab} &= \left< 
\left[\hat\xi(\vecr_a)-\xi(\vecr_a)\right]
\left[\hat\xi(\vecr_b)-\xi(\vecr_b)\right]
\right> \nonumber \\
&= \frac{2}{V}\int \frac{d^3k}{(2\pi)^3} P(\vk)^2 e^{i\vk\cdot(\vecr_a-\vecr_b)},
\end{align}
in other words the Fourier transform of $P^2$ evaluated at $\vecr_a-\vecr_b$.
As we assume $P$ is isotropic, the result depends only on $|\vecr_a-\vecr_b|$.
To compute the covariance between a given vector and the average over
a spherical shell of separation vectors involves averaging this $C_{ab}$
over the spherical shell in $\vecr_b$.  By rotational symmetry, this is 
equivalent to also averaging $\vecr_a$ over a spherical shell.

The key piece of fortune is that if $\xi \propto r^{-2}$,
then $P$ is proportional to $k^{-1}$.  This means that $P^2\propto k^{-2}$,
and the Fourier transform of that is $r^{-1}$.  In other words, we have
\begin{equation}
C_{ab} \propto \frac{1}{|\vecr_a-\vecr_b|}.
\end{equation}
We now need to average this over a shell, but this is a familiar problem,
as this is the same mathematics as the potential of the inverse square
force law.  The solution for spheres is well known from Gauss's law.  
The potential of a sphere is constant inside the sphere (the force being zero)
and then drops as $1/r$ outside the sphere (the force being $1/r^2$).

Hence, we find our first important result that for our stated problem,
the covariance $C_{ab}$ for spheres of radius $r_a$ and $r_b$ is
just proportional to $1/\max(r_a,r_b)$.  
There is a small adjustment to the diagonal elements that goes as 
second order in the shell thickness, computed as the potential 
energy of a shell due to itself.  We neglect this correction in what
follows.

Next, we note that matrices of this form have tridiagonal inverses.
Defining 
\begin{equation}
C = \left(\begin{array}{ccccc}
a_1 & a_2 & a_3 & a_4 & \ldots \\
a_2 & a_2 & a_3 & a_4 & \ldots \\
a_3 & a_3 & a_3 & a_4 & \ldots \\
a_4 & a_4 & a_4 & a_4 & \ldots \\
\vdots & \vdots & \vdots & \vdots & \\
\end{array}\right)
\end{equation}
we write the Fisher matrix as
\begin{equation}
\Phi = C^{-1} = \left(\begin{array}{ccccc}
d_1 & e_1 &  0  &  0  & \ldots \\
e_1 & d_2 & e_2 &  0  & \ldots \\
 0  & e_2 & d_3 & e_3 & \ldots \\
 0  &  0  & e_3 & d_4 & \ldots \\
\vdots & \vdots & \vdots & \vdots & \\
\end{array}\right).
\end{equation}
Solving the linear set of equations, we find
\begin{align}
e_j &= -{1\over a_j-a_{j+1}} \\
d_j &= -e_{j-1} - e_j 
\end{align}
where we define $a_0 = \infty$ and $a_{N+1} = 0$, thereby implying $e_0=0$.

In our application, if we have radii $r_j$ for $j=1\ldots N$, then we
have $a_j = 1/r_j$.  Note that $a_j>a_{j+1}$, so $e_j<0$ and $d_j>0$.
If we simplify to the case of shells spaced evenly with spacing $\Delta$, 
then we have $r_{j+1} = r_j+\Delta$.  We then find $d_j = 2r_j^2/\Delta$
and $e_j = -(r_j^2+r_j\Delta)/\Delta$ for the bulk of the matrix; 
the first and last values are different.
Interestingly, if one then forms the correlation coefficient, one
finds $e_j/\sqrt{d_j d_{j+1}} = -1/2$, again excluding the first and 
last values.  This form is provocative, as it indicates that the Fisher matrix $\Phi = C^{-1}$
is close to the second derivative operator.  This is suggested by the 
fact that $P^{-2} \propto k^2$.

When fitting models, we use this matrix $\Phi$ to compute $\chi^2 =
\vec\delta^T \Phi \vec\delta$, where $\vec\delta = \vec w_{\rm{model}} - \vec w_{\rm{data}}$ is the residual 
between the data and the model correlation function.  For $\xi \propto r^{-2}$, we can factor the Fisher matrix 
\begin{equation}
\Phi_{ij} = \Phi_{ii}^{1/2} R^{ij} \Phi_{jj}^{1/2},
\end{equation}
where $\bf{R}$ is 
the correlation coefficient matrix given by
\begin{equation}
\bf{R} = \left(\begin{array}{ccccc}
1 & -\frac{1}{2} &  0  &  0  & \ldots \\
-\frac{1}{2} & 1 & -\frac{1}{2} &  0  & \ldots \\
 0  &-\frac{1}{2} & 1 & -\frac{1}{2} & \ldots \\
 0  &  0  & -\frac{1}{2} & 1 & \ldots \\
\vdots & \vdots & \vdots & \vdots & \\
\end{array}\right),
\label{equ:R_matrix}
\end{equation}
and $\Phi_{ii}^{1/2} = \sqrt{d_i}$ is a diagonal matrix.

We then want
to consider factorizations $R = K K^T$.  With that, we 
transform $\vec{\delta}$ to $\vec{y}$ with $y_i = K^T_{ij} \Phi{jj}^{1/2} \delta_j=[\Phi^{1/2}]{ij} \delta_j$
and get $\chi^2 = \vec{y}^T \vec{y}$, meaning that we have identified a
set of bins that are statistically independent from one another.

In general, a tridiagonal matrix $T$ has a Cholesky decomposition, in which $T = LL^T$
where $L$ is lower triangular, that itself is zero except for the diagonal
and first subdiagonal.  In the limit that $R$ is a large tridiagonal matrix of 
unit diagonal with $-0.5$ off-diagonal, the Cholesky decomposition well away
from the boundary effects at the end converges to $1/\sqrt2$ on the diagonal
and $-1/\sqrt2$ in the sub-diagonal.  This means that the vector $\vec{y}$
is converging to a first derivative of $\delta$ (after rescaling by the 
square root of the diagonal of $\Phi$).  This is a transformation of the
correlation function that is very compact, only two elements, and yet yields
a set of nearly independent bins.

Alternatively, one can consider the symmetric square root of the
tridiagonal $R$ matrix.  The symmetric square root of a tridiagonal
matrix is no longer tridiagonal, so the resulting independent mode
is not as compact.  The particular $R$ here, with $-0.5$ on the
sub-diagonal, has a symmetric square root that approaches the form
$R^{1/2}_{ij} = 9/10[1-4(i-j)^2]$ when one is far from the edge of
the matrix (i.e., in the large rank limit).  This has a similar
form to $R$, namely positive on the diagonal and negative in the
off-diagonals, but the off-diagonals decay only as $(i-j)^{-2}$
instead of being truncated after the first term.

Clearly the result that the inverse of the covariance matrix is
tridiagonal is linked to the input assumption that the correlation
function is proportional to $r^{-2}$.  Deviations will create a
more extensive matrix.  However, we suggest that the fact that many
galaxy samples do have correlation functions similar to this
particular power-law --- see \citet{2006Masjedi} for a dramatic example
--- is why it turns out that inverse covariance matrices in realistic
cases are found to need only a few off-diagonal terms to describe 
them.  

One can make further use of the result that $C_{ab} \propto |\vec{r}_a
- \vec{r}_b|^{-1}$ to consider the implication for non-spherically
averaged bins.  For example, let us consider the common case in
which the bins inside an annulus are weighted by a Legendre polynomial
of the angle to the line of sight.  In other words, we are considering
the monopole, quadrupole, etc., of the anisotropic correlation
function.  The electrostatic potential resulting from a spherical
harmonic distribution of charge on a shell is simply solved, resulting
in a potential that uses the same spherical harmonic times a mononial
in radius.  The covariance between two such distributions will be
related to $\int d^3r \rho_a \Phi_b$, where $\rho_a$ is the weighting
of the first bin and $\Phi_b$ is the potential resulting from the
weighting of the second bin.  Since $\Phi_b$ preserves the same
spherical harmonic, we find that the answer will be zero if bins
$a$ and $b$ are of different Legendre orders, e.g., we find that
the monopole and quadrupole of the correlation function are
statistically independent.  Further, one could compute the covariance
of the quadrupole at different scales.  We remind that this is only
for the Gaussian random field limit; one expects non-Gaussian terms to
be important at smaller separations.

\section{Simulation and mocks}
\label{sec:simulation}

To showcase our methodology on more realistic correlation functions, we first describe our simulations and mocks. 
We use the \textsc{AbacusCosmos} suite of emulator cosmological simulations generated by the fast and high-precision \textsc{Abacus} N-body code \citep[][Ferrer et al., in preparation; Metchnik $\&$ Pinto, in preparation]{2018Garrison, 2016Garrison}. Specifically we use a series of 16 cyclic boxes with Planck 2016 cosmology \citep{2016Planck} at redshift $z = 0.5$, where each box is of size 1100~$h^{-1}$~Mpc, and contains 1440$^3$ dark matter particles of mass $4\times 10^{10}$ $h^{-1}M_{\odot}$. The force softening length is 0.06~$h^{-1}$~Mpc. Dark matter halos are found and characterized using the ROCKSTAR \citep{2013Behroozi} halo finder. 

We generate mock Luminous Red Galaxy (LRG) catalogs using a standard 5-parameter Halo Occupation Distribution (HOD) model \citep{2007bZheng, 2009Zheng, 2015Kwan}. We also incorporate redshift-space distortions (RSD) effects. The details of our implementation can be found in \citet{2017Yuan, 2018Yuan}.

Given a mock catalog, the anisotropic 2PCF is estimated using the SciPy kD-tree based pair-counting routine,
\begin{equation}
\xi(d_{\perp}, d_{\parallel}) = \frac{N_{\rm{mock}}(d_{\perp}, d_{\parallel})}{N_{\rm{rand}}(d_{\perp}, d_{\parallel})} - 1,
\label{equ:xi_def}
\end{equation}
where $d_{\parallel}$ and $d_{\perp}$ are the projected separation along the line-of-sight (LOS) and perpendicular to the LOS respectively. $N_{\rm{mock}}(d_{\perp}, d_{\parallel})$ is the number of galaxy pairs within each bin in $(d_{\perp}, d_{\parallel})$, and $N_{\rm{rand}}(d_{\perp}, d_{\parallel})$ is the expected number of pairs within the same bin but from a uniform distribution.  

The projected 2PCF $w(d_{\perp})$ is then estimated as 
\begin{equation}
w(d_{\perp}) = \int_{-d_{\parallel, \rm{max}}}^{d_{\parallel, \rm{max}}} \xi(d_{\perp}, d_{\parallel}) d(d_{\parallel}),
\label{equ:w_def}
\end{equation}
where $d_{\parallel, \rm{max}} = \sqrt{d_{p, \rm{max}}^2 - d_{\perp}^2}$ is the maximum separation along the LOS given $d_{\perp}$ and a maximum separation $d_{p,\rm{max}}$. For the rest of this paper, we choose $d_{p,\rm{max}} = 30$~Mpc. We choose a uniform binning of 30 bins between $0 < d_{\perp} < 10$~Mpc. Figure~1 of \citet{2018Yuan} illustrates the anisotropic 2PCF and the projected 2PCF computed from our mock catalogs. 

\section{Decorrelating the projected 2PCF}
\label{sec:projected} 

In this section, we use the mock galaxy catalogs to propose linear transformation matrices for the projected 2PCF that largely decorrelate the separation bins and produce nearly diagonal covariance matrices.

Using the mocks, we first compute the covariance matrix of the projected 2PCF, $C(w_i, w_j)$, where $i,j = 1,2,...,30$ denote the bin numbers. We divide each of the sixteen $1100 h^{-1}$~Mpc boxes into 125 equal sub-volumes, yielding a total of 2000 sub-volumes and a total volume of $21.3 h^{-3}$~Gpc$^3$. We calculate the covariance matrix from the dispersion among the 2000 sub-volumes. The degree of division is chosen to have a large number of sub-volumes, yet ensuring that each sub-volume is large enough that the dispersion among the sub-volumes is not dominated by sample variance. 

In practice, we prewhiten the projected 2PCF computed in each sub-volume with the mean projected 2PCF $\bar{w}(d_{\perp})$, computed across all 16 boxes. Specifically, the prewhitened projected 2PCF is defined as $\hat{w}(d_{\perp}) = w(d_{\perp})/\bar{w}(d_{\perp})$. The idea of prewhitening in the context of correlation functions is not new \citep{2000aHamilton, 2000bHamilton}, and it produces a flatter covariance matrix with a well-behaved inverse. For the rest of this paper, we study the covariance matrix of the prewhitened projected 2PCF, $C(\hat{w}_i, \hat{w}_j)$, which we simply denote as $C_{ij}(\hat{w})$. 

\begin{figure*}
\centering
 \hspace*{-0.4cm}
\includegraphics[width=6in]{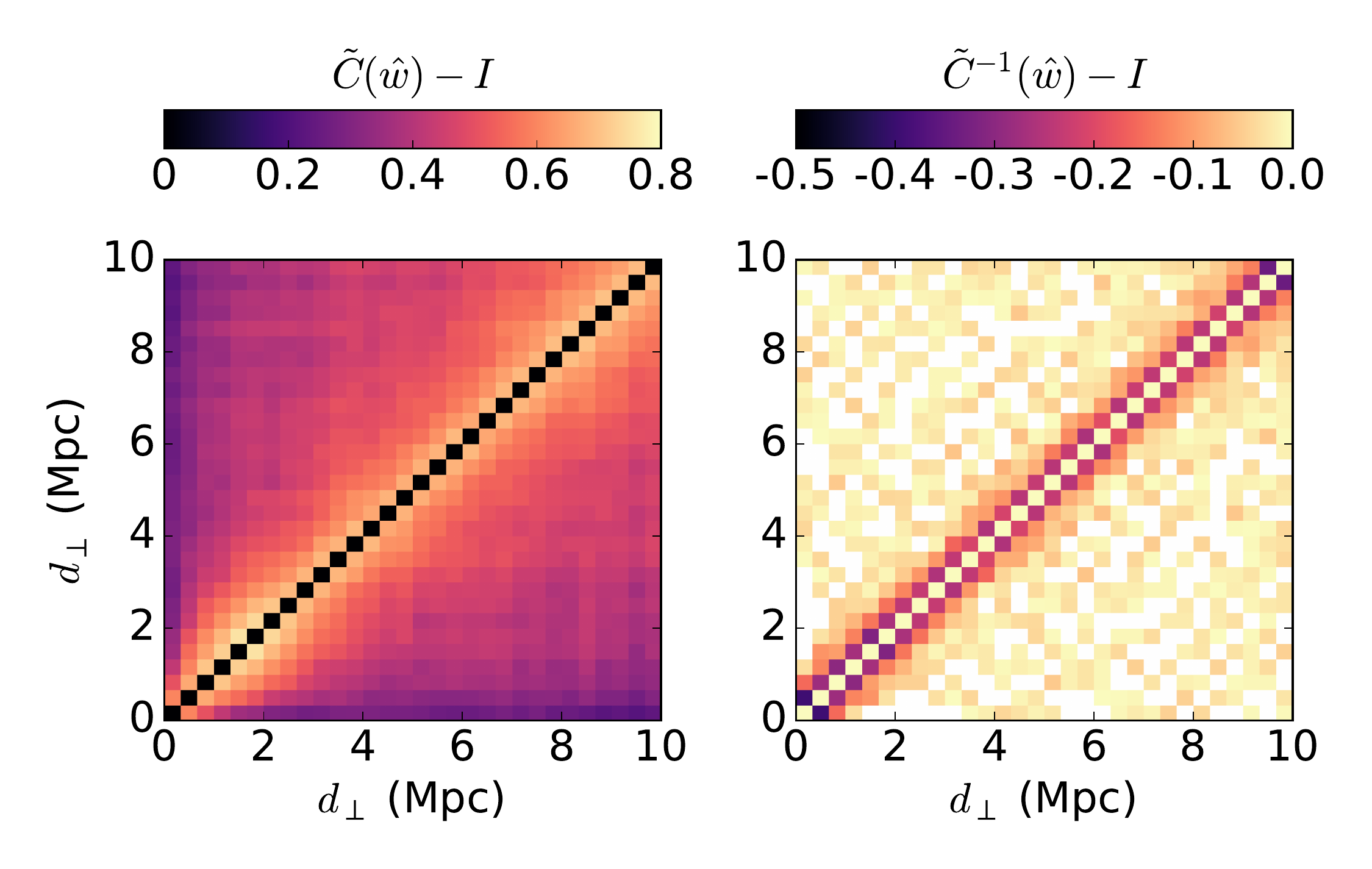}
\vspace{-0.4cm}
\caption{The left panel shows the reduced covariance matrix of the prewhitened projected 2PCF $\hat{w}$, with the diagonal subtracted off to reveal off-diagonal structures. The diagonal is subtracted off to reveal the structure of the off-diagonals. The right panel shows the inverse covariance matrix, again normalized to show the correlation coefficients. }
\label{fig:cov_dw_normed}
\end{figure*}

To facilitate comparison between covariance matrices, we define the reduced covariance matrix as 
\begin{equation}
\tilde{C}_{ij} = \frac{C_{ij}}{\sqrt{C_{ii}C_{jj}}}.
\label{equ:normalization}
\end{equation}
By construction, the reduced covariance matrix has unity on the diagonal. We subtract off the identity matrix to showcase the off-diagonal terms of the reduced covariance matrix in Figure~\ref{fig:cov_dw_normed}. If the bins of the 2PCF are perfectly decorrelated, then we expect the reduced covariance matrix to be the identity matrix, and the $\tilde{C} - I$ matrix to be 0. The goal of this section is thus to find simple transformation matrices on the 2PCF that minimize the values of the $\tilde{C} - I$ matrix.
We use the tilde on top notation to denote the reduced covariance matrix throughout this paper. 
The left panel of Figure~\ref{fig:cov_dw_normed} shows the reduced covariance matrix of the prewhitened projected 2PCF $\hat{w}$. We see strong covariance ($\tilde{C}_{ij}(\hat{w}) > 0.5$) between bins less than ${\sim} 2$~Mpc apart, while the covariance weakens moderately towards smaller $d_\perp$. 

To quantitatively compare the covariance matrices as we introduce transformations that aim to minimize the values of the $\tilde{C} - I$ matrix, we define the mean residual of the covariance matrix to be the mean of the absolute values of all the off-diagonal terms of the reduced covariance matrix, barring the edge rows and columns. We remove the edge rows and columns as they suffer from edge effects in the transformation. The mean residual will be used as a measure of the residual correlation between the 2PCF bins after applying the transformation matrix. For reference, the reduced covariance matrix of the projected 2PCF with no transformation (left panel of Figure~\ref{fig:cov_dw_normed}) has a mean residual of 0.489.


To test the performance of our models in a realistic scenario, we also include a second quantitative measure that we call $\chi^2$. In an ideal situation where we know the true covariance matrix $C$, the $\chi^2$ would simply be given by 
\begin{equation}
\chi^2_{\rm{True}} = \delta \hat{w}^T C^{-1}\delta \hat{w},
\label{equ:chi2_true}
\end{equation}
where $\delta\hat{w}$ is some arbitrary perturbation to the projected 2PCF. However, when we do not have access to the full covariance matrix, for example on a plot only showing the diagonal error bars, we can estimate the $\chi^2$ to be 
\begin{equation}
    \chi^2 \approx \sum_i (\delta \hat{w}_i/\sigma_i)^2,
    \label{equ:chi2_estimat}
\end{equation}
where $i$ is the index of the separation bins and $\sigma_i^2$ are the diagonal elements of the covariance matrix. Our claim is that our transformation matrices produce nearly diagonal covariance matrices, so that the diagonal elements of the transformed covariance matrix do capture most of errors. Thus, if we estimate the $\chi^2$ with transformed 2PCF and the diagonal of the transformed covariance matrix, the estimated $\chi^2$ will be close to $\chi^2_{\rm{True}}$. 
To illustrate this point, we use an arbitrary HOD perturbation to induce a change to the projected 2PCF, which we use throughout this paper when estimating $\chi^2$.
To generate an example perturbation to the 2PCF $\delta \hat{w}$, we employ an HOD perturbation that can be found in the first row of Table~1 of \citet{2018Yuan}. The resulting $\chi^2_{\rm{True}} = 17.77.$ If we do not use any transformations and estimate $\chi^2$ with just the diagonal of the covariance matrix, we get $\chi^2 = 76.66$. With our transformation matrices, we aim to bring the estimated $\chi^2$ closer to $\chi^2_{\rm{True}}$. We point out that we use the first row of Table~1 of \citet{2018Yuan} as our $\delta \hat{w}$ in all quoted $\chi^2$ values throughout this paper, but we repeat the test with other HOD perturbations as well.

The right panel of Figure~\ref{fig:cov_dw_normed} shows the reduced form of the inverse covariance matrix, with the identity subtracted off. The inverse covariance is compact around the diagonal, with most of the power in the first two off-diagonals. This is broadly consistent with our prediction for $\xi \propto r^{-2}$. However, the value on the first off-diagonal is approximately $-0.25$, whereas Equation~\ref{equ:R_matrix} predicts the value of the first off-diagonal to be -0.5. We also see some residual power at farther off-diagonals. These are signs that the monopole 2PCF does not exactly follow $\xi \propto r^{-2}$. Thus, we expect our subsequent transform matrices, specifically the Cholesky matrix and the symmetric square root, will be broader than for $\xi \propto r^{-2}$. 

In the rest of this section, we use our mock galaxy catalogs to develop compact models of the transformation matrix based on the Cholesky decomposition and the symmetric square root of the inverse covariance matrix that produce largely decorrelated separation bins. We present best fits for these models and use them to transform the projected 2PCF and its covariance matrix. We compare these models with the metrics we just described and identify one that best decorrelates the projected 2PCF and produces a nearly diagonal covariance matrix.

\subsection{Cholesky Decomposition}
\label{subsec:projected_cholesky} 

As we have described in Section~\ref{sec:theory}, the Cholesky decomposition of the inverse covariance matrix provides a compact transformation matrix that decorrelates the $1/r^2$ 2PCF. In the first part of this section, we showed a case where the anisotropic 2PCF does not scale exactly as $1/r^2$. The corresponding inverse covariance matrix $\Phi$ is not exactly tridiagonal, which in turn means that the Cholesky decomposition of $\Phi$ has some residual power beyond the first off-diagonal.

\begin{figure}
\centering
 \hspace*{-0.4cm}
\includegraphics[width=3.7in]{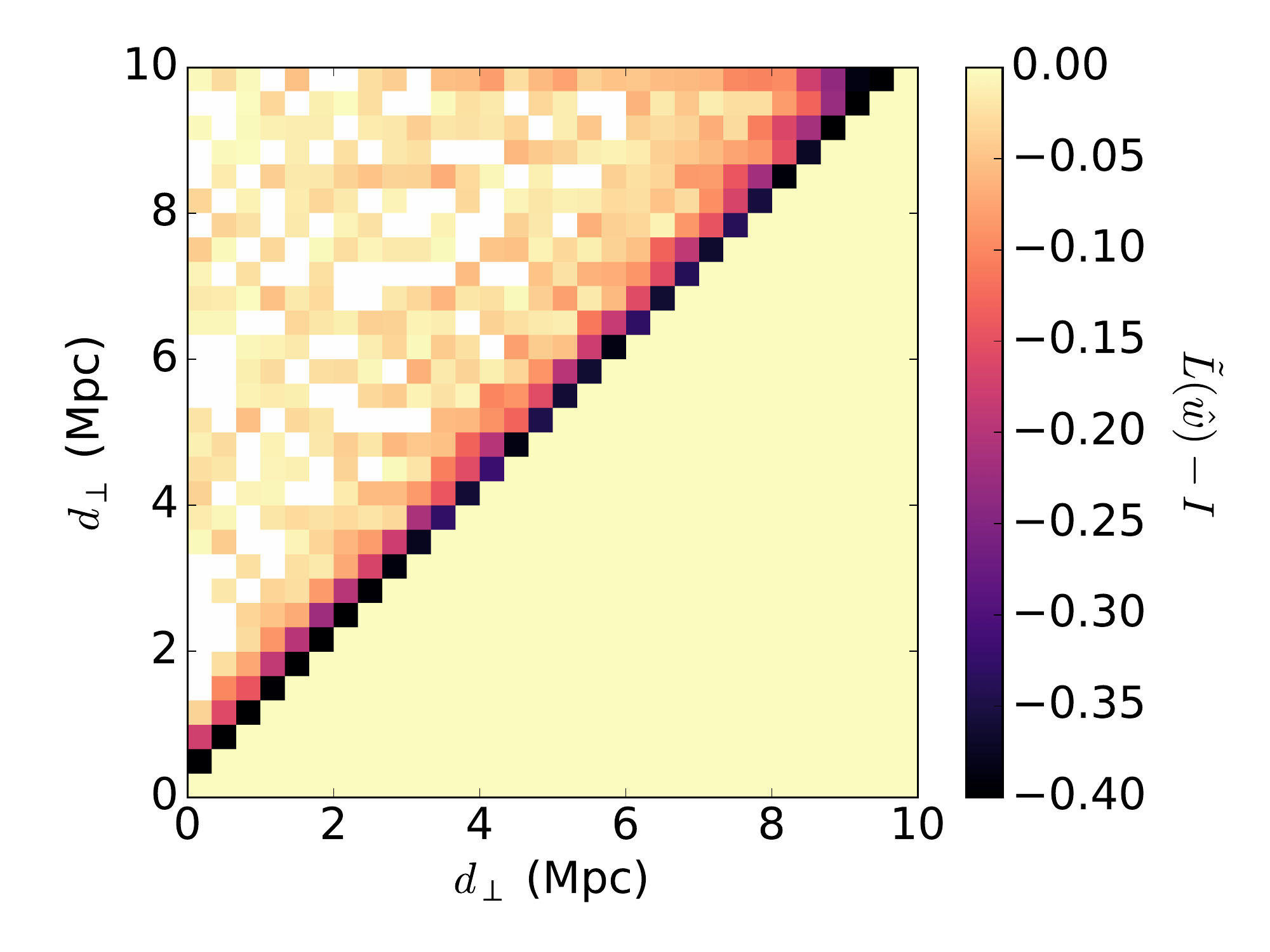}
\vspace{-0.4cm}
\caption{The reduced Cholesky matrix $\tilde{L}$, where $\Phi_{ij} = LL^T$. The normalization is similar to Equation~\ref{equ:normalization} to give unity on the diagonal. The diagonal is then subtracted to reveal the structure of the off-diagonals. }
\label{fig:cholesky_dw_normed}
\end{figure}

Figure~\ref{fig:cholesky_dw_normed} shows the reduced Cholesky matrix $\tilde{L}$, where $\Phi_{ij} = LL^T$. We see that the Cholesky matrix is compact, with strong signal on the first and second off-diagonals and some residuals on the further off-diagonals. The mean of the first off-diagonal is approximately 0.4, off from the prediction of $1/\sqrt{2}\approx 0.7$ for the $\xi \propto r^{-2}$ case. This difference again emphasizes the fact that the 2PCF of the mock galaxies does not follow exactly a inverse square law, but that the scaling is sufficiently close to an inverse square law that we recover a compact Cholesky matrix. 

We first follow a similar routine to that of \citet{2014Anderson} to construct a compact transformation matrix that approximates the Cholesky matrix. We start with a upper triangular matrix that is unity on the diagonal and non-zero on only the first two off-diagonals. We choose the first and second diagonal to be uniform and equal to the mean of the first and second off-diagonal of the Cholesky matrix, respectively. The transformed projected 2PCF is given by
\begin{equation}
\hat{w}_{\rm{Ch}, i} = \frac{\hat{w}_i + a\hat{w}_{i-1} + b\hat{w}_{i-2}}{1+a+b}.
\label{equ:cholesky_transform}
\end{equation}
$\hat{w}_{\rm{Ch}, i}$ and $\hat{w}_i$ are the transformed and pre-transform projected 2PCF in the $i$-th bin, respectively. The values of $a$ and $b$ are case specific and sensitive to the amount of sample variance and shot noise. A fair choice is to take the mean of the first and second off-diagonals of the Cholesky matrix, respectively. For our mocks, we get $a \approx -0.387, b \approx -0.181$. We have added in the normalization to preserve the sum of the 2PCF over all bins. 

\begin{figure*}
\centering
 \hspace*{-0.4cm}
\includegraphics[width=6in]{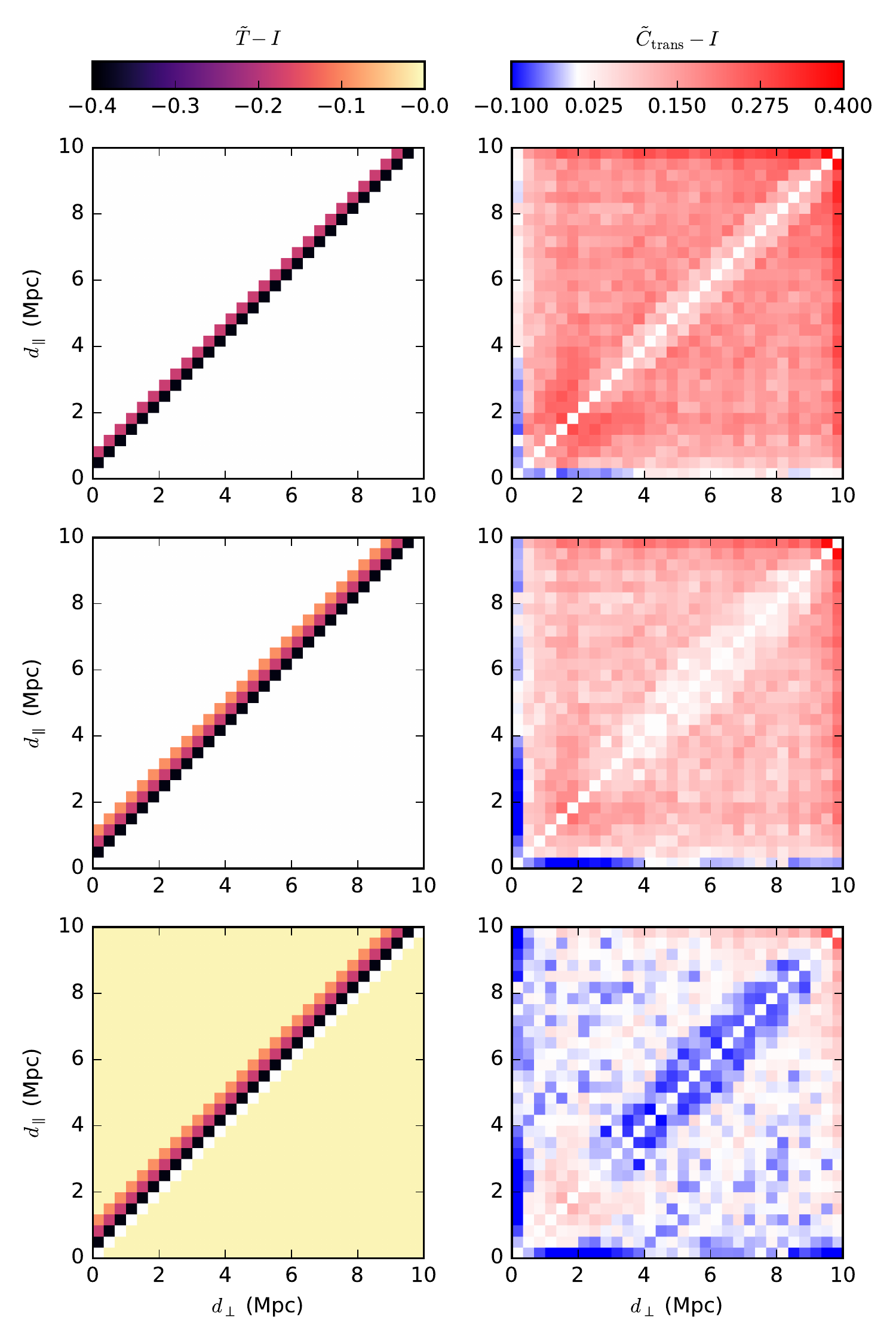}
\vspace{-0.4cm}
\caption{The left panels show 3 different transformation matrices based on the Cholesky matrix. From top to bottom, the three rows correspond to Equation~\ref{equ:cholesky_transform}-\ref{equ:cholesky_longer_transform}, respectively. The matrix is normalized with the diagonal and then we subtract off the identity matrix to focus on the off-diagonal terms. The right panels show the reduced covariance matrix of the transformed projected 2PCF $\hat{w}_{\rm{Ch}}$ using the corresponding transformation matrix. Again, the unity diagonal has been subtracted off to reveal the structure of the off-diagonals.}
\label{fig:cov_cholesky}
\end{figure*}

The top left panel of Figure~\ref{fig:cov_cholesky} shows the resulting transform matrix, whereas the top right panel shows the reduced covariance matrix of the transformed projected 2PCF $\hat{w}_{\rm{Ch}}$. The subscript ``trans" stands for ``transformed," which we take to denote the covariance matrix of the transformed 2PCF. The tilde again denotes the reduced covariance matrix defined by Equation~\ref{equ:normalization}. We subtract off the identity matrix to showcase the off-diagonal terms of the reduced covariance matrix.
Comparing to the pre-transform covariance matrix shown in the left panel of  Figure~\ref{fig:cov_dw_normed}, we see that our simple transformation has greatly suppressed the off-diagonal covariances. The mean residual of the transformed covariance matrix is 0.154, which represents a ${\sim} 68\%$ suppression compared to the mean residual of 0.489 in the pre-transform case. This is impressive considering that our transformation matrix is compact in scale and is only parameterized by 2 parameters. Using the diagonal of the transformed covariance matrix and following Equation~\ref{equ:chi2_estimat}, we estimate $\chi^2 = 36.71$, which is much closer to $\chi^2_{\rm{True}}$ compared to the pre-transform $\chi^2 = 76.66$.

To combat the remaining residual signal in the off-diagonals of the transformed covariance matrix, we adopt two modifications to our transform matrix. First, we include a third uniform off-diagonal, with the value set to the mean of the third off-diagonal of the Cholesky matrix. This transformation can be expressed as
\begin{align}
\hat{w}'_{\rm{Ch}, i} = \frac{1}{N}[&\hat{w}_i + a \hat{w}_{i-1} + b \hat{w}_{i-2} + c \hat{w}_{i-3}],
\label{equ:cholesky_long_transform}
\end{align}
where $a \approx -0.387, b \approx -0.181, c \approx -0.092$ and $N = 1 + a + b + c$ is the normalization constant. The corresponding transformation matrix and the transformed covariance matrix are shown in the second row of Figure~\ref{fig:cov_cholesky}. We see a further suppression to the off-diagonal terms of the transformed covariance matrix. The corresponding mean residual is 0.094, which represents an ${\sim} 81\%$ reduction compared to the pre-transform case. Following Equation~\ref{equ:chi2_estimat}, the estimated $\chi^2 = 29.37$, an improvement compared to the 2-diagonal model. 

The second modification we apply is to add a pedestal value $\epsilon$ to the full transformation matrix,
\begin{align}
\hat{w}''_{\rm{Ch}, i} = \frac{1}{N}[&(1-\epsilon)\hat{w}_i + (a-\epsilon) \hat{w}_{i-1} + (b-\epsilon) \hat{w}_{i-2} \nonumber \\ &+ (c-\epsilon) \hat{w}_{i-3} + \epsilon\sum_{k = 1}^{30}\hat{w}_{k}],
\label{equ:cholesky_longer_transform}
\end{align}
where $N$ is the normalization constant, and $a$, $b$, and $c$ are the same constants as in Equation~\ref{equ:cholesky_long_transform}. The pedestal value $\epsilon$ is fitted to minimize the square sums of the off-diagonal terms in the transformed covariance matrix. For our mocks, we have $\epsilon \approx -0.007$.

The final transformation matrix with the best fit $\epsilon$ is plotted in the bottom left panel of Figure~\ref{fig:cov_cholesky}. The first three off-diagonals are the same as the middle left panel. The $\epsilon$ term is added to the rest of the matrix, both in the upper and lower triangular half, seen as the uniform background color in the figure. Note the diagonal is kept at exactly zero.

The bottom right panel of Figure~\ref{fig:cov_cholesky} shows the reduced covariance matrix of the transformed projected 2PCF $\hat{w}''_{\rm{Ch}}$. Comparing to the covariance matrices in the top and middle row, we see that we have successfully suppressed the far off-diagonal terms to very close to 0. Ignoring the edges, the only notable residuals left are in the region of $d_\perp \sim 2$~Mpc and a slightly negative trough in the first few off-diagonals. The corresponding mean residual of the transformed covariance matrix is 0.026, representing a ${\sim} 95\%$ suppression compared to the pre-transform case. The estimated $\chi^2 = 16.58$, very close to the true value $\chi^2_{\rm{True}} = 17.77$.
These results show that by introducing a compact and simple 4-parameter ($a,b,c,\epsilon$) model of the Cholesky decomposition of the Fisher matrix, we have successfully reduced the the correlations between the projected 2PCF bins by ${\sim} 95\%$.

\begin{figure}
\centering
 \hspace*{-0.4cm}
\includegraphics[width=3.5in]{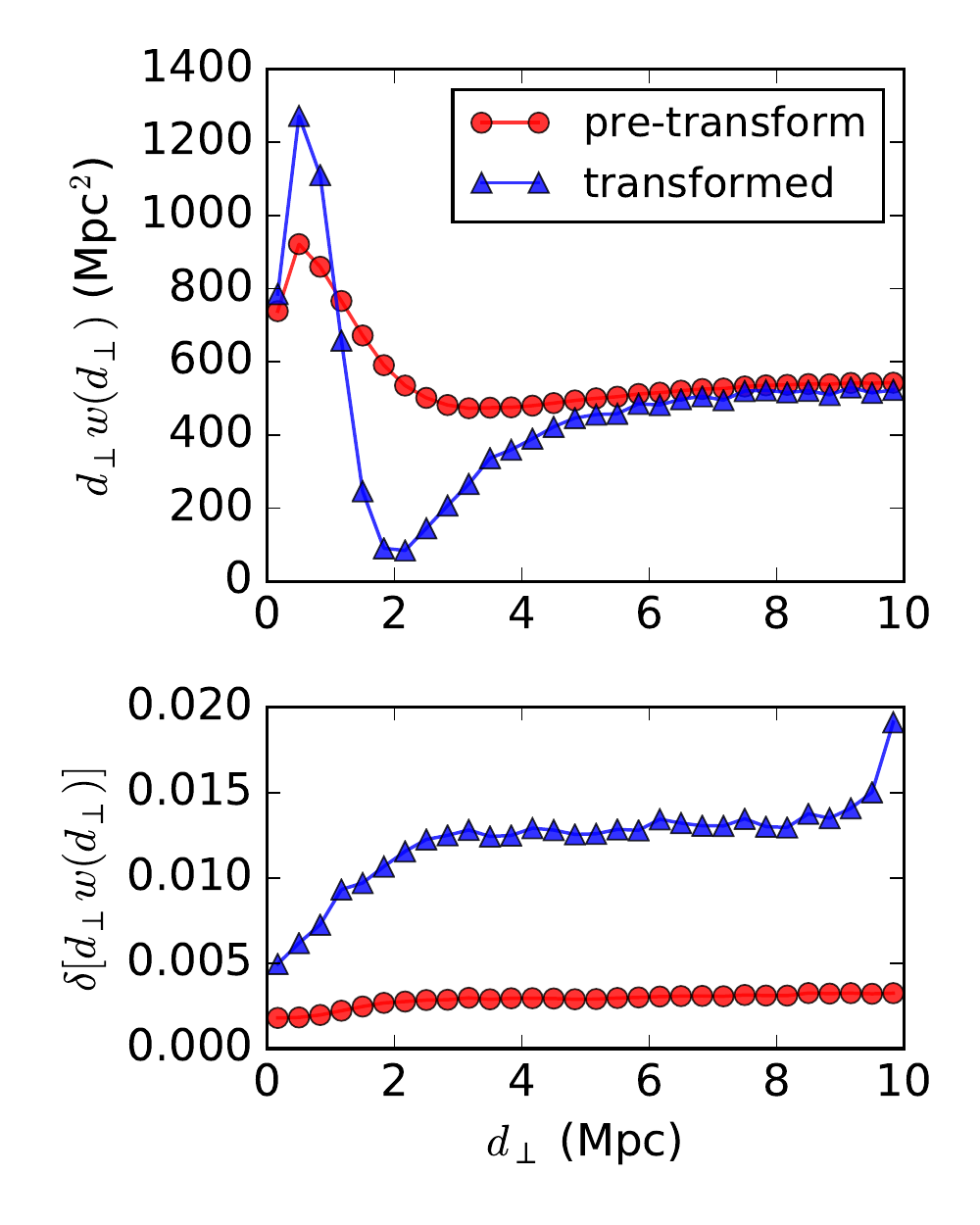}
\vspace{-0.4cm}
\caption{The top panel shows the pre-transform projected 2PCF in red and the transformed projected 2PCF in blue, weighted by $d_\perp$. The transform is parametrized with 3 uniform off-diagonals and a broad pedestal value (Equation~\ref{equ:cholesky_longer_transform}) and shown in its matrix form in the bottom left panel of Figure~\ref{fig:cov_cholesky}. The bottom panel shows the relative error bars on each of the bins for the pre-transform and transformed projected 2PCF. The error bars are computed as $1\sigma$ deviation around the mean and then normalized by $d_\perp w(d_{\perp})$. We see much larger error bars in the transformed 2PCF.}
\label{fig:2pcf_smoothed_cholesky}
\end{figure}

The top panel of Figure~\ref{fig:2pcf_smoothed_cholesky} shows the transformed projected 2PCF $d_\perp w''_{\rm{Ch}}$ in blue and the pre-transform projected 2PCF $d_\perp w$ in red. We see that the transformed 2PCF recovers the same qualitative features as the pre-transform 2PCF. Specifically, we recover the maximum at around $0.5$~Mpc and the minimum at 2 to 3 Mpc, albeit the pre-transform minimum is at larger $d_{\perp}$ and is much less pronounced. Both curves also start at around the same value and eventually flatten out to approximately the same value at large separation. The bottom panel shows the $1\sigma$ error bars on the transformed 2PCF and the pre-transform 2PCF in relative units. The error bars are computed from the diagonals of the covariance matrices, $C_{ij}(\hat{w})$ and $C_{ij}(\hat{w}''_{\rm{Ch, i}})$, respectively. We see that while the transformed 2PCF has similar amplitude as the pre-transform 2PCF, the error bars on the transformed 2PCF are more than $100\%$ larger. This is because the error bars of the pre-transform 2PCF underestimate the uncertainties by ignoring the off-diagonal terms of the covariance matrix, whereas the covariance matrix of the transformed 2PCF (bottom right panel of Figure~\ref{fig:cov_cholesky}) is mostly diagonal. Thus, the transformed 2PCF has error bars much more representative of the level of uncertainties in the statistics. 

\subsection{Fisher Square root}
\label{subsec:projected_fisher} 

A second way to model the transformation matrix is to model it on the symmetric square root of the Fisher matrix, which we simply refer to as the Fisher square root from now on.

\begin{figure}
\centering
 \hspace*{-0.4cm}
\includegraphics[width=3.7in]{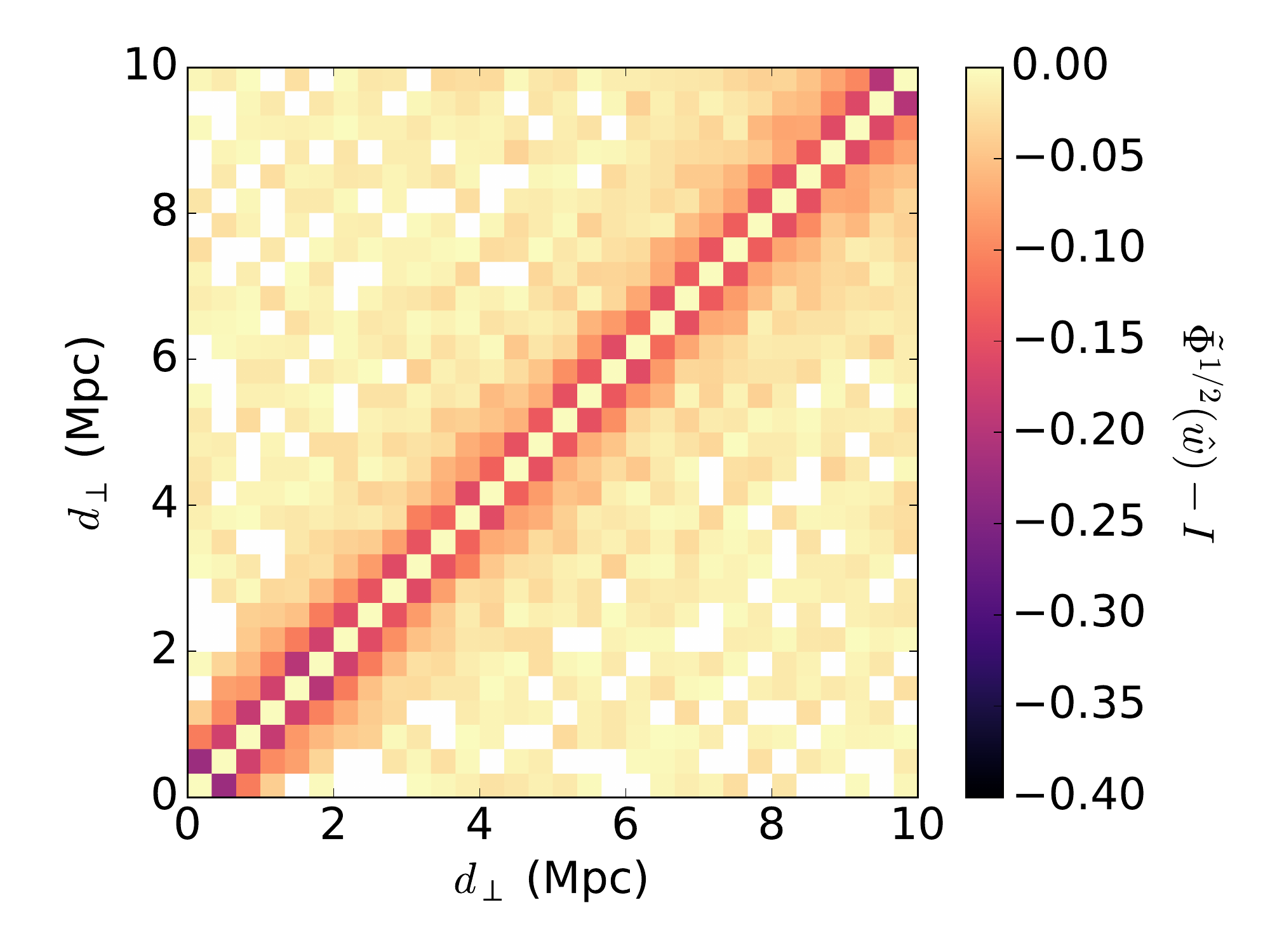}
\vspace{-0.4cm}
\caption{The reduced symmetric square root of the Fisher matrix $\tilde{\Phi}^{1/2}$. The normalization is similar to that of Equation~\ref{equ:normalization}. The diagonal is then subtracted to reveal the structure of the off-diagonals. }
\label{fig:invsqrt_dw_normed}
\end{figure}

Figure~\ref{fig:invsqrt_dw_normed} shows the reduced Fisher square root $\tilde{\Phi}^{1/2}$. By definition, this matrix is the transformation matrix that would fully decorrelate the projected 2PCF. We see a strong band structure, with high power in the first few off-diagonals. This suggests that applying a relatively narrow transformation kernel can largely decorrelate the projected 2PCF and suppress the off-diagonal terms of the covariance matrix.

Following the same procedure as for the Cholesky case, we reproduce the pentadiagonal transformation matrix given by 
\begin{equation}
\hat{w}_{\rm{FS}, i} = \frac{\hat{w}_i + a(\hat{w}_{i-1}+\hat{w}_{i+1}) + b(\hat{w}_{i-2}+\hat{w}_{i+2})}{1+2a+2b}.
\label{equ:penta_transform}
\end{equation}
where $\hat{w}_i$ is the prewhitened projected 2PCF in the $i$-th bin, and $\hat{w}_{\rm{FS}, i}$ is the corresponding transformed 2PCF in the same bin. The denominator serves as a normalization constant. The only unknown parameters of such a pentadiagonal transform matrix are the weights on the first and second off-diagonals, $a$ and $b$. We set the weights $a$ and $b$ to the mean of the first and second off-diagonals of the normalized Fisher square root. For our mocks, we get $a \approx -0.156$ and $b \approx -0.086$. 

\begin{figure*}
\centering
 \hspace*{-0.4cm}
\includegraphics[width=6in]{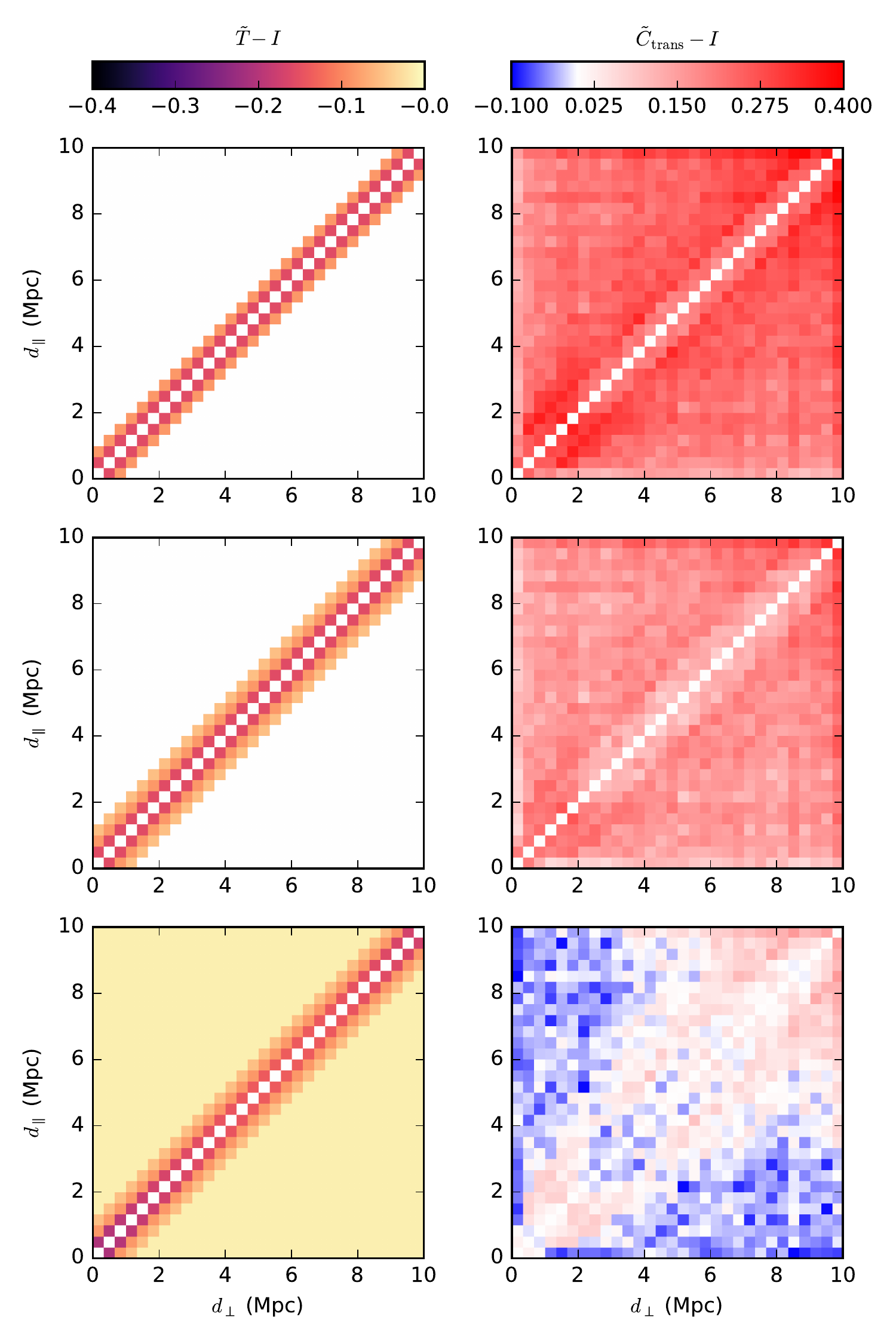}
\vspace{-0.4cm}
\caption{The left panels show 3 different transformation matrices based on the Fisher square root matrix. From top to bottow, the three rows correspond to Equation~\ref{equ:penta_transform}-\ref{equ:penta_transform_longer}, respectively. The matrices are normalized with the diagonal and then we subtract off the identity matrix to focus on the off-diagonal terms. The right panels show the reduced covariance matrix of the transformed projected 2PCF $\hat{w}_{\rm{FS}}$ using the corresponding transformation matrix. Again, the unity diagonal has been subtracted off to reveal the structure of the off-diagonals.}
\label{fig:cov_invsqrt}
\end{figure*}

The top left panel of Figure~\ref{fig:cov_invsqrt} shows the matrix form of the pentadiagonal transformation matrix given by Equation~\ref{equ:penta_transform}. The top right panel shows the reduced covariance matrix of the transformed projected 2PCF $\hat{w}_{\rm{FS}}$. Comparing to the covariance of the pre-transform 2PCF shown in Figure~\ref{fig:cov_dw_normed}, we see strong suppression to the off-diagonal terms of the covariance matrix. The mean residual is 0.236, which is $52\%$ suppression compared to the mean residul of the pre-transform covariance matrix. However, we still have strong residual covariances in the third and further off-diagonals, at ${\sim} 20-40\%$. We also see a strong scale dependence of the residual covariance, with strong residuals at the smallest scales $< 3$~Mpc and at larger scale $> 8$~Mpc. With this transform, we follow Equation~\ref{equ:chi2_estimat} and estimate the $\chi^2 = 45.45$, a substantial improvement over the pre-transform estimate of 76.66.

Similar to what we did with the Cholesky case, we apply three modifications to the transform matrix to suppress the residual covariances. First, we include a third off-diagonal in the transformation matrix. Again we take the value of the third off-diagonal to be the mean of the third off-diagonal of the Fisher square root matrix. The resulting transformation is given by
\begin{equation}
\hat{w}'_{\rm{FS}, i} = \frac{\hat{w}_i + a(\hat{w}_{i-1}+\hat{w}_{i+1}) + b(\hat{w}_{i-2}+\hat{w}_{i+2}) + c(\hat{w}_{i-3}+\hat{w}_{i+3})}{1+2a+2b+2c}.
\label{equ:penta_transform_long}
\end{equation}
where the notations are defined similarly as Equation~\ref{equ:penta_transform}. $c$ is the value of the third off-diagonal. For our mocks, we have $a \approx -0.156, b\approx -0.086, c\approx -0.053$. 
The corresponding transform matrix is shown in the middle left panel of Figure~\ref{fig:cov_invsqrt}, and the corresponding transformed covariance matrix is shown in the middle right panel. The mean residual is now 0.163, which represents a ${\sim} 67\%$ reduction compared to the pre-transform case. However, there is still significant residual in the first three off-diagonals at the very small scale ${\sim} 2$~Mpc, and the broad residual in the far off-diagonals persist. The estimated $\chi^2 = 36.81$, a moderate improvement compared to the 2-diagonal model.

To combat the residual in the first three off-diagonals, we include scale dependence in the first off-diagonal of the transformation matrix. This is also motivated by the strong scale dependence seen in the first off-diagonal of the Fisher square root as shown in Figure~\ref{fig:invsqrt_dw_normed}, where the first off-diagonal shows more negative signal towards both the smaller scale and the larger scale. We model the first off-diagonal of the transformation matrix as a quadratic function where all three parameters are fitted using least-squares to the two first off-diagonals shown in Figure~\ref{fig:invsqrt_dw_normed}. 

To suppress the broad residual covariances, we adopt the same approach as we did for the Cholesky case. We introduce a pedestal value $\epsilon$ to the transform matrix and fit it to minimize the square sums of the off-diagonal terms of the transformed covariance matrix. 
Thus, we propose an alternative transformation kernel of the following form,
\begin{align}
\hat{w}''_{\rm{FS}, i} = \frac{1}{N}[&(1-\epsilon)\hat{w}_i + (a_i-\epsilon)(\hat{w}_{i-1}+\hat{w}_{i+1}) + (b-\epsilon)(\hat{w}_{i-2}+\hat{w}_{i+2}) \nonumber \\ &+ (c-\epsilon)(\hat{w}_{i-3}+\hat{w}_{i+3}) + \epsilon\sum_{k = 1}^{30}\hat{w}_{k}],
\label{equ:penta_transform_longer}
\end{align}
where the contribution from the first off-diagonal $a_i$ now depends on scale. Parameter $b$ and $c$ are again the mean of the second and third off-diagonal of the Fisher square root respectively. The pedestal value $\epsilon$ is fitted to minimize the square sum of the off-diagonal terms of the transformed covariance matrix. For our mocks, we have $b\approx -0.086, c\approx -0.053, \epsilon \approx -0.011$. The first diagonal is parametrized by a quadratic model of the form $A + B(x - C)^2$, where $x$ is the bin number. The best fit values are $A = -0.14, B = -2.5\times 10^{-4}$, and $C = 16$.

The bottom left panel of Figure~\ref{fig:cov_invsqrt} shows the transformation matrix described by Equation~\ref{equ:penta_transform_longer}. Note the color gradient along the first off-diagonal showing the quadratic fit. The pedestal value is reflected in the uniform color in the far off-diagonals. The corresponding transformed covariance matrix is shown in the bottom right panel. Comparing to the top right panel, we see that the largest off-diagonal now is ${\sim} 0.1$ instead of ${\sim} 0.3$, barring the edges. The mean residual is $0.031$, which corresponds to a ${\sim} 94\%$ reduction in off-diagonal covariances compared to the pre-transform case. Using this transformation matrix, we estimate $\chi^2 = 20.93$, close to the true value of 17.77.

Thus, we have constructed a simple and compact transformation matrix based on the Fisher square root using 6 parameters (3 for the quadratic fit of the first off-diagonal, and $b,c,\epsilon$). As a result, we have successfully reduced the correlations between the projected 2PCF bins by ${\sim} 94\%$, allowing for a simple model of the Fisher matrix and accurate representations of errors. 

\begin{figure}
\centering
 \hspace*{-0.4cm}
\includegraphics[width=3.5in]{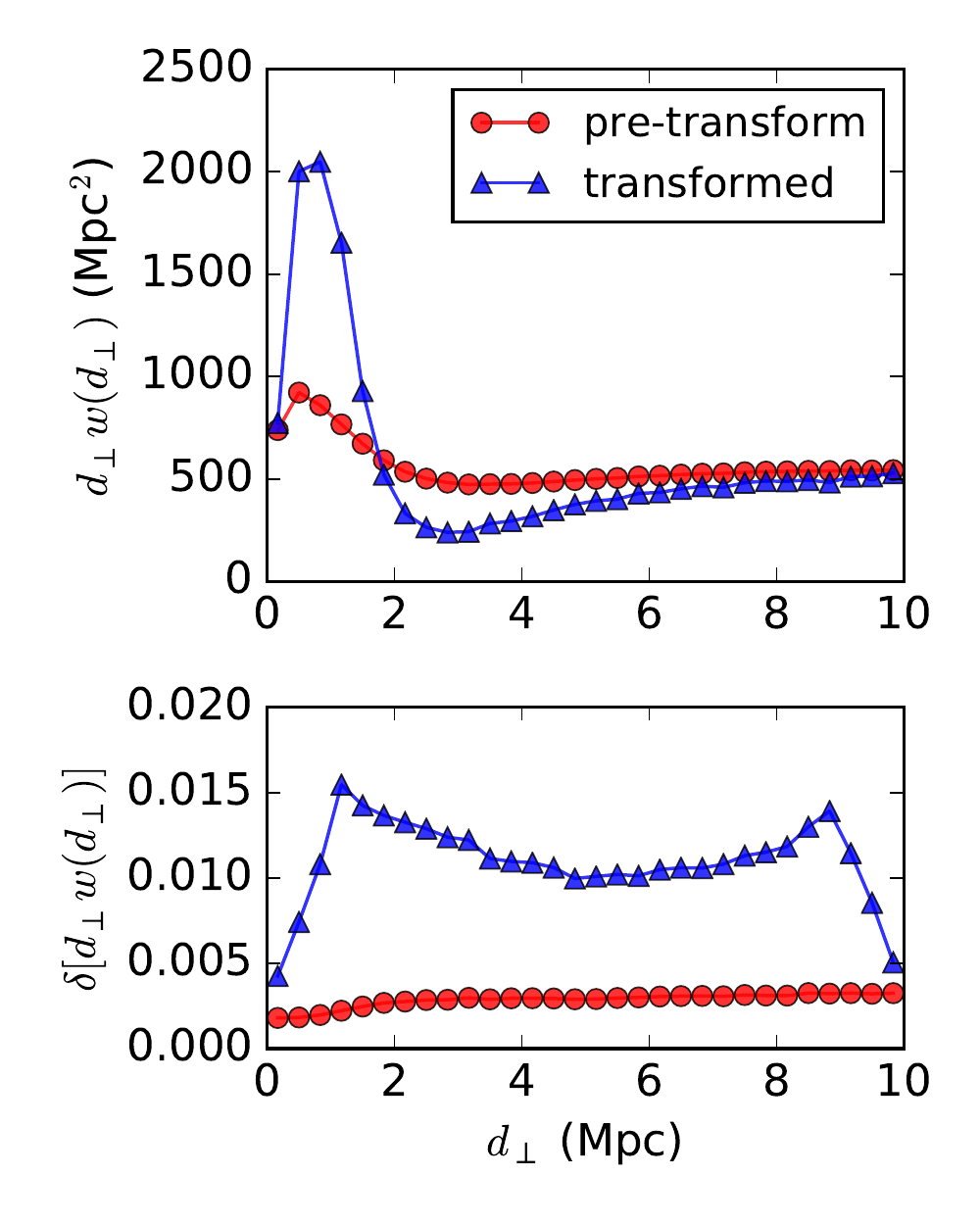}
\vspace{-0.4cm}
\caption{The top panel shows the pre-transform projected 2PCF in red and the transformed projected 2PCF in blue, weighted by $d_\perp$. The transform is parametrized with 3 uniform off-diagonals and a broad pedestal value (Equation~\ref{equ:penta_transform_longer}) and shown in its matrix form in the bottom left panel of Figure~\ref{fig:cov_invsqrt}. The bottom panel shows the relative error bars on each of the bins for the pre-transform and transformed projected 2PCF. The error bars are computed as $1\sigma$ deviation around the mean and then normalized by $d_\perp w(d_{\perp})$. We see larger error bars in the transformed 2PCF.}
\label{fig:2pcf_smoothed_invsqrt}
\end{figure}

The top panel of Figure~\ref{fig:2pcf_smoothed_invsqrt} shows the pre-transform projected 2PCF in red and the transformed 2PCF in blue, weighted by $d_\perp$. The transform is described by Equation~\ref{equ:penta_transform_longer}. We again see that the transformed 2PCF recovers the same qualitative features as the pre-transform 2PCF, specifically in the maximum, the minimum, and the flat tail at large scale. The relative error bars shown in the bottom panel correspond to one standard deviation around the mean, which are computed from the diagonal terms of the covariance matrices $C_{ij}(\hat{w})$ and $C_{ij}(\hat{w}''_{\rm{FS}})$. Due to our suppression of the off-diagonal terms of the covariance matrix, the bins of the transformed projected 2PCF are largely decorrelated. The error bars of the transformed 2PCF more accurately capture the uncertainties in the statistics, whereas the error bars of the pre-transform 2PCF underestimate the uncertainties.


\begin{table}
\begin{tabular}{ c | c c c}
\hhline {====}
 & $N_{\rm{params}}$ & mean residual & $\chi^2$ \\ 
\hline
Cholesky 2 (Eq. \ref{equ:cholesky_transform})  & 2 & 0.154 & 36.71 \\ 

Cholesky 3 (Eq. \ref{equ:cholesky_long_transform}) & 3 & 0.094 & 29.37 \\ 

Cholesky 3+$\epsilon$ (Eq. \ref{equ:cholesky_longer_transform}) & 4 & 0.026 & 16.58 \\
\hline 
FS 2 (Eq. \ref{equ:penta_transform}) & 2 & 0.236 & 45.45 \\ 

FS 3 (Eq. \ref{equ:penta_transform_long}) & 3 & 0.163 & 36.81 \\ 

FS 3+$\epsilon$ (Eq. \ref{equ:penta_transform_longer}) & 6 & 0.031 & 20.93 \\
\hline 
\end{tabular} 
\caption{A summary of all the transformations we proposed for the projected 2PCF. The first column lists the names of the transformations and the corresponding equation number. ``FS" stands for Fisher square root, and the number following describes the number of off-diagonals used. $\epsilon$ signals the use of a pedestal value. The second column and the third column summarizes the number of parameters used to construct the transformation matrix and the resulting mean residual in the transformed covariance matrix. The fourth column lists the estimated $\chi^2$ of the transformed 2PCFs with just the diagonal of the transformed covariance matrix (Equation~\ref{equ:chi2_estimat}). The 2PCF perturbation is drawn from the HOD perturbation quoted in the first row of Table~1 of \citet{2018Yuan}. The pre-transform covariance matrix has a mean residual of 0.489 for reference. The true $\chi^2_{\rm{True}} = 17.77$.}
\label{tab:summary_wp}
\end{table}

We summarize our transformation matrices and their corresponding transformed covariance matrices in Table~\ref{tab:summary_wp}. The first column shows the names of transformations and their corresponding equation numbers. The second column shows the number of parameters needed to construct the transformation matrices. The third column shows the mean residual of the transformed covariance matrices. The fourth column lists the estimated $\chi^2$ following Equation~\ref{equ:chi2_estimat} using just the diagonal of the transformed covariance matrices. Comparing the mean residuals, we see that a moderate increase in the number of parameters in the transformation matrix leads to a dramatic decrease in the mean residual. The inclusion of a small broad pedestal value $\epsilon$ proves to be critical in reducing the mean residual to ${\sim} 5\%$ of the pre-transform case. We see the same trend in the estimated $\chi^2$ values. Cross-comparing the mean residuals and the $\chi^2$ between the Cholesky cases and the FS cases, we also see that the Cholesky based transformation matrices consistently outperform the Fisher square root cases. Especially when we include 3 diagonals and a pedestal value, the Cholesky based transformation matrix outperforms the Fisher square root case while requiring 2 fewer parameters. We have validated these results with several different HOD perturbations.

\section{Decorrelating the anisotropic 2PCF}
\label{sec:anisotropic}
In this section, we extend our methodology to the anisotropic 2PCF. 
We compute the covariance matrix of the anisotropic 2PCF $\xi$ in the same way as we did for the projected 2PCF. We divide each of the 16 simulation boxes into 125 equal sub-volumes and compute the covariances in $\xi$ from the dispersion among all the sub-volumes. 
Since $\xi$ is strongly dependent on scale, we calculate the fractional anistropic 2PCF $\hat{\xi} = \xi/\bar{\xi}$ (pre-whitening), where $\xi$ is computed in each sub-volume and $\bar{\xi}$ is the average across all sub-volumes. This prewhitens the covariance matrix of $\hat{\xi}$, which ensures more stable matrix inversions and optimizations. 

The anisotropic 2PCF is binned in ($d_\perp, d_\parallel$) plane, specifically with 15 bins between 0 and 10~Mpc along the $d_\perp$ axis and 9 bins between 0 and 27~Mpc along the $d_\parallel$ axis. The bins are then flattened into a 1D array of length 135 in a column-by-column fashion, where each column corresponds to 9 bins along the $d_\parallel$ axis.

\begin{figure}
\centering
 \hspace*{-0.4cm}
\includegraphics[width=3.7in]{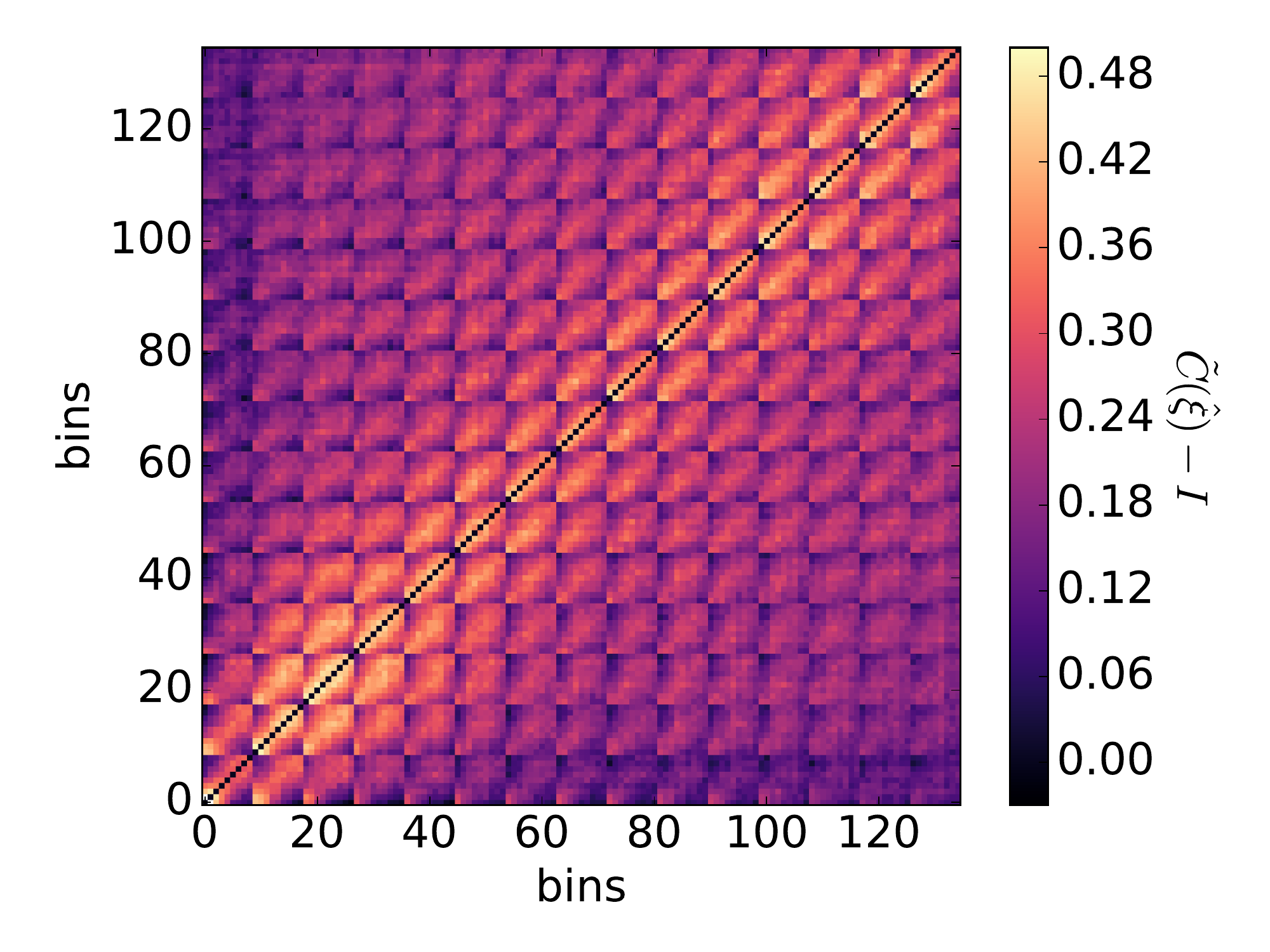}
\vspace{-0.4cm}
\caption{The reduced covariance matrix of the prewhitened anisotropic 2PCF, $\tilde{C}(\bar{\xi}_i, \bar{\xi}_j)$. The bins are iterated along $d_\parallel$, i.e. each 9 bin block corresponds to the same $d_\perp$.}
\label{fig:cov_xi}
\end{figure}

Figure~\ref{fig:cov_xi} shows the 135$\times$135 reduced covariance matrix with the diagonal subtracted off. We see high off-diagonal power in the covariance matrix, especially around $d_\perp\sim 2$~Mpc. There is also a strong periodic band structure as we go farther off the diagonal. Each band represents the covariance of a bin with bins in the neighboring columns. In addition, each diagonal is also no longer uniform, but rather shows a periodic pattern along the diagonal every 9 bins. This second periodic pattern is due to the flattening of the $\hat{\xi}$ matrix into an array. When the bins wrap around into the next or previous column, then the covariances drop off, as these bins are far apart in the $(d_\perp, d_\parallel)$ space. These periodic band structures persist in the following subsections in our models of the Cholesky decomposition and symmetric square root of the Fisher matrix. 

Similar to Section~\ref{sec:projected}, we define the mean residual of a covariance matrix to be the mean of the absolute value of the off-diagonal terms of the reduced covariance matrix. For our mocks, the mean residual of the pre-transform covariance matrix of the anistropic 2PCF is 0.221.
We also perform the same $\chi^2$ test as we have done for the projected 2PCFs. We adopt the same HOD perturbation, quoted in the first row of Table 1 of \citet{2018Yuan}. Using the resulting perturbations in the anisotropic 2PCF and the full covariance matrix, we get the true $\chi^2_{\rm{True}} = 1234$. If we only use the diagonal of the covariance matrix instead, we get an estimated $\chi^2 = 2206$.
 
\subsection{Cholesky decomposition}

\begin{figure}
\centering
 \hspace*{-0.8cm}
\includegraphics[width=3.7in]{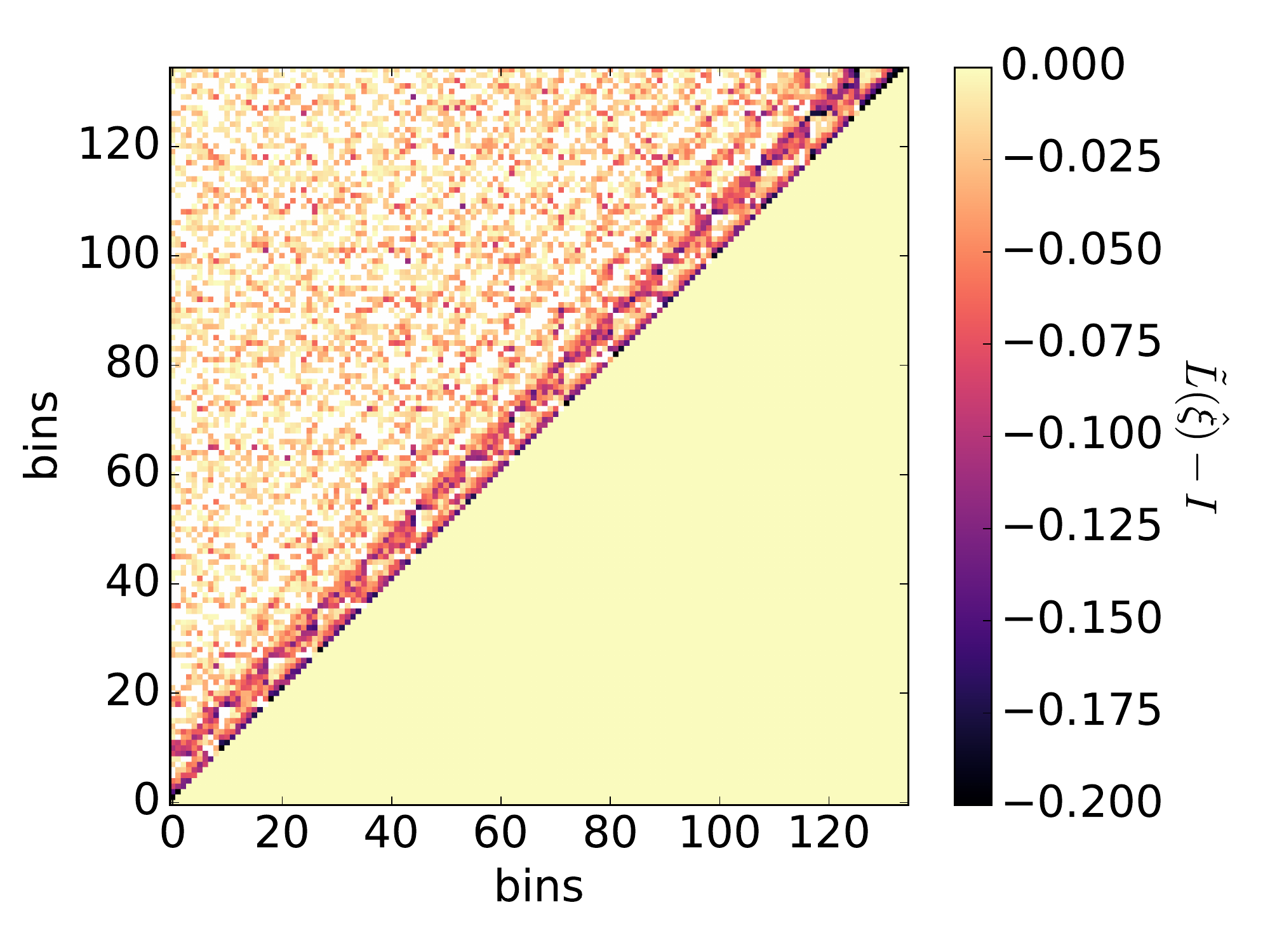}
\vspace{-0.4cm}
\caption{The Cholesky decomposition of the inverse covariance matrix for the prewhitened anisotropic 2PCF $\hat{\xi}$. The matrix is normalized and the diagonal is subtracted off to reveal the off-diagonals. We see a strong band structure that propagates to far off-diagonals.}
\label{fig:cholesky_xi_normed}
\end{figure}

We first model a transformation matrix on the Cholesky decomposition of the inverse covariance matrix. We show the Cholesky matrix for the prewhitened anisotropic 2PCF in Figure~\ref{fig:cholesky_xi_normed}. We have again normalized the matrix with the diagonal and then subtracted off the identity to reveal the off-diagonal structure. We see both periodic band structures in the covariance matrix persists in the Cholesky matrix. 

To model the transformation on the Cholesky matrix, we construct a compact kernel in $(d_\perp, d_\parallel)$ space. This is analogous to what we did with the projected 2PCF, where we constructed a compact kernel in $d_\perp$ space that extends only to the closes $2-3$ bins in each direction, before we added in the broad residual terms. Here, we first construct a transform kernel that spans 2 bins in both directions along the $d_\perp$ axis and 2 bins in both directions along the $d_\parallel$ axis. We visualize this kernel as 
\begin{equation}
\centering
\begin{array}{*{20}c}
           &       &\times &       &  \\
           &\times &\times &0      &  \\ 
    \times &\times &1      &0      &0  \\ 
           &\times &0      &0      &  \\ 
           &       &0      &       &  \\
 \end{array},
 \label{equ:kernel_2}
\end{equation}
where the horizontal axis correspond to the $d_\perp$ axis and the vertical axis corresponds to the $d_\parallel$ axis. Each $\times$ symbolizes a kernel weight to be determined, and we have normalized the kernel so that the center value is 1. The zeros are placed to ensure that the corresponding transformation matrix is upper triangular. 

\begin{figure*}
\centering
 \hspace*{-0.4cm}
\includegraphics[width=6in]{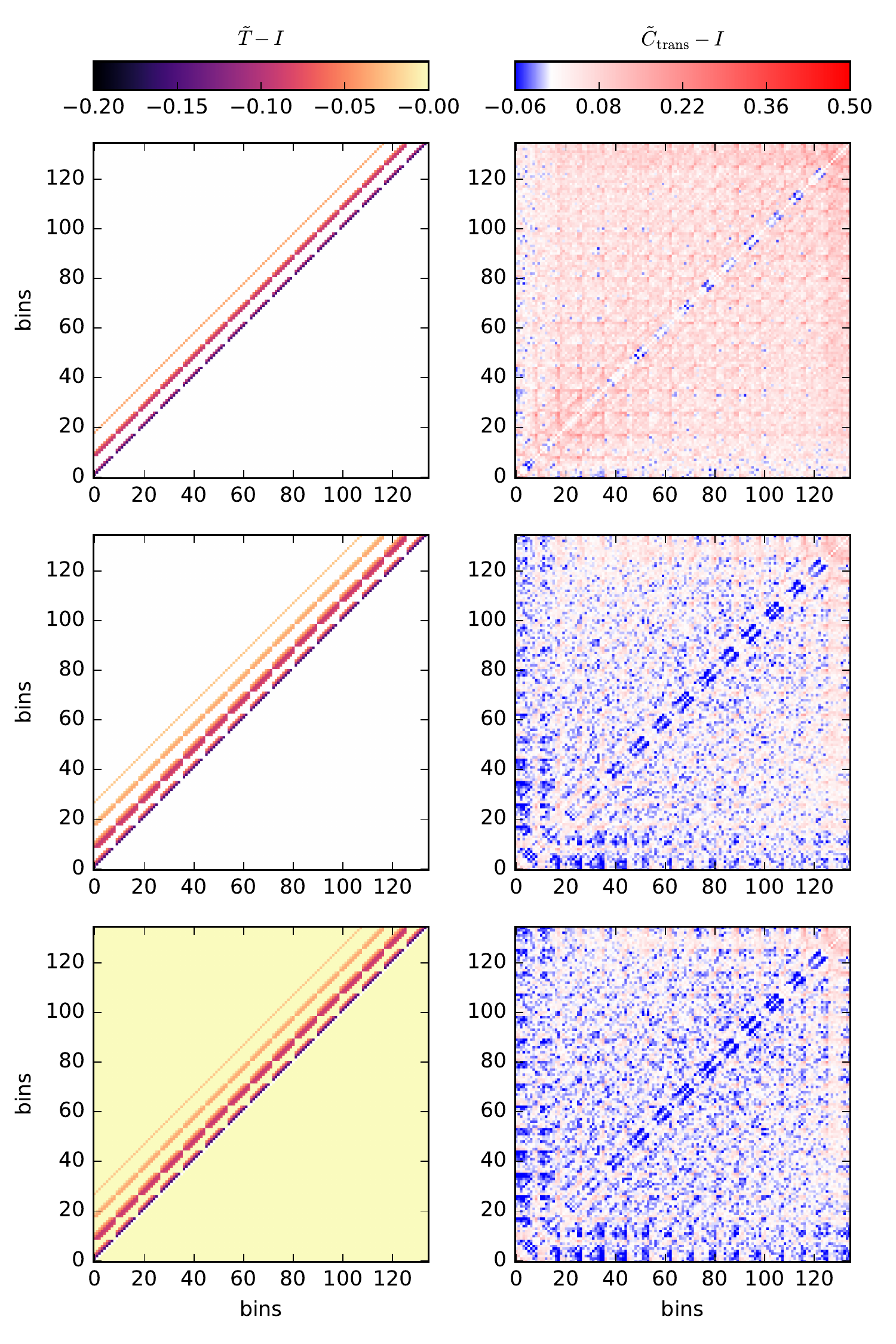}
\vspace{-0.4cm}
\caption{The left panels show 3 different approximations to the Cholesky matrix for the prewhitened anisotropic 2PCF. The matrix is normalized with the diagonal and then we subtract off the identity matrix to reveal the off-diagonal terms. The top and middle row correspond to Equation~\ref{equ:kernel_2_vals} and Equation~\ref{equ:kernel_3_vals}, respectively. The bottom row adds a small pedestal value to the full transformation matrix. The right panels show the reduced covariance matrix of the transformed anisotropic 2PCF after applying the corresponding transformation matrix. Again, the unity diagonal has been subtracted off to reveal the structure of the off-diagonals.}
\label{fig:cov_cholesky_xi}
\end{figure*}

The corresponding transformation matrix will have a triple band structure, with each column in Equation~\ref{equ:kernel_2} occupying one band. The structure of the transformation matrix is visualized in the top left panel of Figure~\ref{fig:cov_cholesky_xi}, where we see the first band of width 2 (diagonal subtracted off), the second band of width 3, and the third band of width 1. We have maintained the periodic structure along each band to prevent the kernel (Equation~\ref{equ:kernel_2}) from crossing the edges in $(d_\perp, d_\parallel)$ space and wrapping around. The kernel weights are computed again by averaging the values of the Cholesky matrix along each off-diagonal. We present these values in Equation~\ref{equ:kernel_2_vals}, which is set up the same way as Equation~\ref{equ:kernel_2}.
\begin{equation}
\centering
\begin{array}{*{20}c}
           &       &-0.096 &       &  \\
           &-0.055 &-0.153 &0      &  \\ 
    -0.035 &-0.088 &1      &0      &0  \\ 
           &-0.092 &0      &0      &  \\ 
           &       &0      &       &  \\
 \end{array},
 \label{equ:kernel_2_vals}
\end{equation}
The corresponding reduced covariance matrix of the transformed anisotropic 2PCF is shown in the top right panel. We see a dramatic suppression to the off-diagonals terms, compared with Figure~\ref{fig:cov_xi}. There is still some residual towards $d_\perp \sim 2$~Mpc, and a broad residual signal across the whole matrix. The mean residual of the transformed covariance matrix is 0.061, representing a ${\sim} 72\%$ suppression relative to the pre-transform case. With this transform, we follow Equation~\ref{equ:chi2_estimat} and estimate the $\chi^2 = 1253$, a substantial improvement over the pre-transform estimate of 2206. 

We follow the same modifications as we did for the projected 2PCF case. We first expand the transformation kernel to include 3 bins in each direction. The resulting kernel can be visualized as 
\begin{equation}
\centering
\begin{array}{*{20}c}
          &       &       &\times &       &       &  \\
          &       &\times &\times &0      &       &  \\ 
          &\times &\times &\times &0      &0      &  \\ 
   \times &\times &\times &1      &0      &0      &0 \\ 
          &\times &\times &0      &0      &0      &  \\ 
          &       &\times &0      &0      &       &  \\ 
          &       &       &0      &       &       &  \\
 \end{array}.
 \label{equ:kernel_3}
\end{equation}
Again the horizontal axis represents the $d_\perp$ direction and the vertical axis represents the $d_\parallel$ direction. We compute the kernel weights by taking the mean of the Cholesky matrix along each off-diagonal. We present these values in Equation~\ref{equ:kernel_3_vals}.
\begin{equation}
\centering
\begin{array}{*{20}c}
          &       &       &-0.054 &       &       &  \\
          &       &-0.034 &-0.096 &0      &       &  \\ 
          &-0.020 &-0.055 &-0.153 &0      &0      &  \\ 
   -0.022 &-0.035 &-0.088 &1      &0      &0      &0 \\ 
          &-0.033 &-0.092 &0      &0      &0      &  \\ 
          &       &-0.080 &0      &0      &       &  \\ 
          &       &       &0      &       &       &  \\
 \end{array}.
 \label{equ:kernel_3_vals}
\end{equation}
The resulting transformation matrix is shown in the middle left panel of Figure~\ref{fig:cov_cholesky_xi}. The corresponding transformed covariance matrix is shown in the middle right panel. We see that this modification has managed to further suppress the residual off-diagonal values compared to the top right panel. Specifically, the residuals in the off-diagonal region around $d_\perp \sim 2$~Mpc are now greatly suppressed compared to the top right panel. The mean residual is 0.029, an approximately $87\%$ suppression compared the pre-transform case. With this transform, we estimate $\chi^2 = 1100$. Note that this $\chi^2$ estimate is actually not as good as the 2-diagonal model, where we got $\chi^2 = 1253$, compared to $\chi^2_{\rm{True}} = 1234$. This turns out to be a special case, and our tests with other HODs show the 3-diagonal model can produce $\chi^2$ estimates that are closer to $\chi^2_{\rm{True}}$ than the 2-diagonal model.

We then introduce one further modification where we add a uniform pedestal value to the whole transformation matrix and fit the value to minimize the square sums of the off-diagonal values of the transformed covariance matrix. The kernel weights shown in Equation~\ref{equ:kernel_3_vals} are held fixed, and the best fit pedestal value is $-3.06\times 10^{-4}$. The resulting transformation matrix is shown in the bottom left panel of Figure~\ref{fig:cov_cholesky_xi}. The pedestal value is too small to be noticeable in the color gradient. The bottom right panel shows the resulting transformed covariance matrix. We see that the introduction of the pedestal value has mildly suppressed the residual covariances between large $d_\parallel$ bins. The mean residual is 0.028, a slight improvement over the the case without the broad pedestal value. The final mean residual represents an ${\sim} 87\%$ suppression in the off-diagonal terms compared to the pre-transform covariance matrix. With this transform, we estimate $\chi^2 = 1119$, a slight improvement over the previous model without a small pedestal term. Compared to $\chi^2_{\rm{True}} = 1234$, the estimated $\chi^2$ is approximately $9\%$ lower than but a dramatic improvement over the pre-transform estimate of 2206.

\begin{figure*}
\centering
 \hspace*{-0.4cm}
\includegraphics[width=6in]{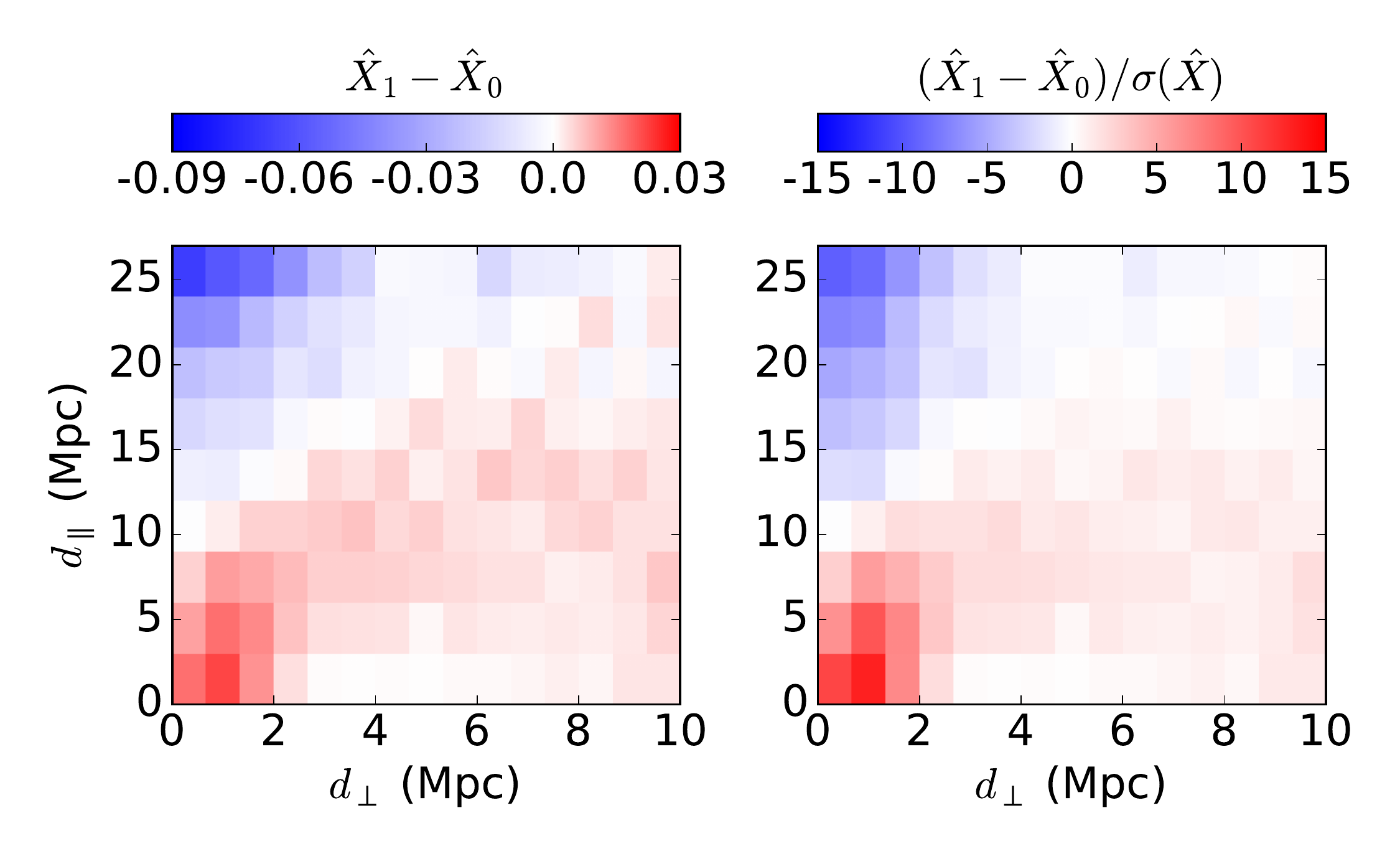}
\vspace{-0.4cm}
\caption{The left panel shows the perturbation to the prewhitened anisotropic 2PCF when we change the HOD from the baseline HOD to the one given by the first row of Table~1 in \citet{2018Yuan}. Compared to the perturbation to the pre-transform anisotropic 2PCF shown in Figure~7 of \citet{2018Yuan}, we see that we have recovered qualitatively the same behavior. The right panel shows the corresponding signal-to-noise in each bin, where the noise is computed from the diagonal of the transformed covariance matrix. We can estimate $\chi^2$ by adding up all the bins on the right panel in quadrature.}
\label{fig:xi_smooth_cholesky}
\end{figure*}

Now that we have constructed a linear transformation that largely decorrelates the prewhitened anisotropic 2PCF $\hat{\xi}$, we would like to showcase the transformed anisotropic 2PCF $\hat{X} = T\hat{\xi}$. By definition, $\hat{X}$ would be a flat unity array since we have pre-whitened $\xi$ and $T$ is normalized to preserve the sum of the anisotropic 2PCF in all the bins.
Instead of showing $\hat{X}$ directly, we show a change in $\hat{X}$ in Figure~\ref{fig:xi_smooth_cholesky}, where we perturb the HOD to introduce a perturbation signal $\hat{X}_1 - \hat{X}_0$ to $\hat{X}$. $\hat{X}_1$ corresponds to the HOD specified in the first row of Table~1 of \citet{2018Yuan}, and $\hat{X}_0$ corresponds to the baseline HOD given by \citet{2007bZheng, 2009Zheng}. The HOD for $\hat{X}_1$ is chosen arbitrarily to show interesting structure along the $d_\perp$ direction. Such structure largely comes from increasing the satellite distribution parameter ($s$, explained in Section~3.2 of \citet{2018Yuan}), which moves satellite galaxies into outer orbits around dark matter halos, biasing their LOS velocity distribution and producing a large LOS signal in the anisotropic 2PCF through RSD.


The left panel of Figure~\ref{fig:xi_smooth_cholesky} shows the perturbation to the transformed 2PCF $\hat{X}$. Compared to the perturbation to the pre-transform anisotropic 2PCF shown in Figure~7 of \citet{2018Yuan}, we see that we have recovered qualitatively the same behavior despite moderate mixing of the scales.  

The right panel shows the corresponding signal-to-noise of this perturbation, where the noise $\sigma(\hat{X})$ are computed from the diagonal of the transformed covariance matrix. We can estimate $\chi^2$ by adding up all the bins on the right panel in quadrature. Thus, one advantage of showing the transformed anisotropic 2PCF is that it gives a much more accurate visual representation of the $\chi^2$ while still preserving the structure of the pre-transform 2PCF. Again we emphasize that this transformation does not affect the actual fitting as one would always use the full covariance matrix anyway.

\subsection{Fisher square root}

\begin{figure}
\centering
 \hspace*{-0.4cm}
\includegraphics[width=3.7in]{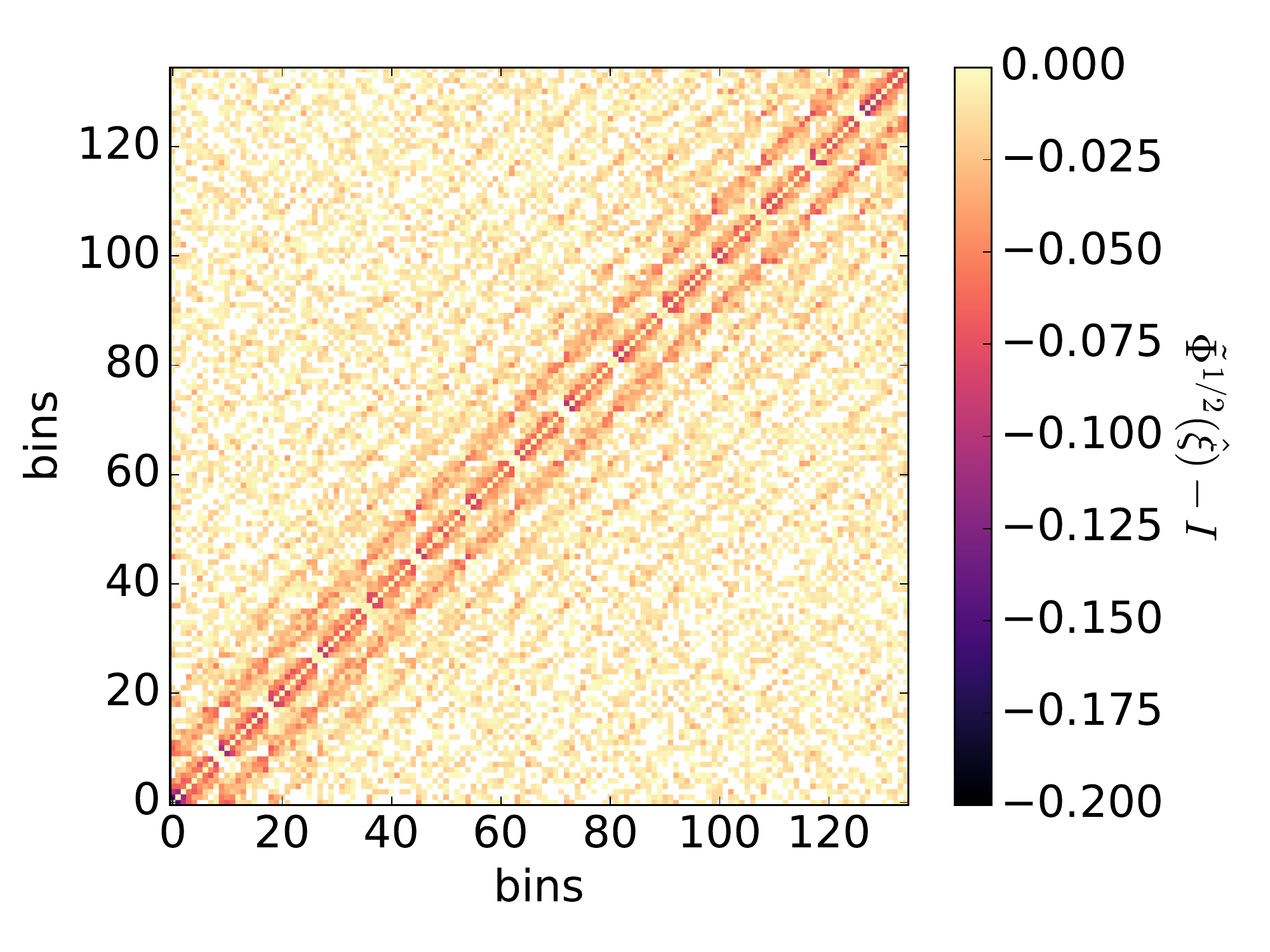}
\vspace{-0.4cm}
\caption{The symmetric square root of the inverse covariance matrix for the prewhitened anisotropic 2PCF $\hat{\xi}$. The matrix is reduced following Equation~\ref{equ:normalization} and the diagonal is subtracted off to reveal the off-diagonals. Similar to the Cholesky matrix, we see a strong band structure that propagates to far off-diagonals and a periodic pattern along the diagonals.}
\label{fig:invsqrt_xi_normed}
\end{figure}

As in the previous section, we now model the transformation matrix on the Fisher square root of the anistropic 2PCF, which we show in Figure~\ref{fig:invsqrt_xi_normed}. Similar to the Cholesky case, we see a periodic band structure that propagates to far off-diagonals and a periodic pattern along each off-diagonal. 

\begin{figure*}
\centering
 \hspace*{-0.4cm}
\includegraphics[width=6in]{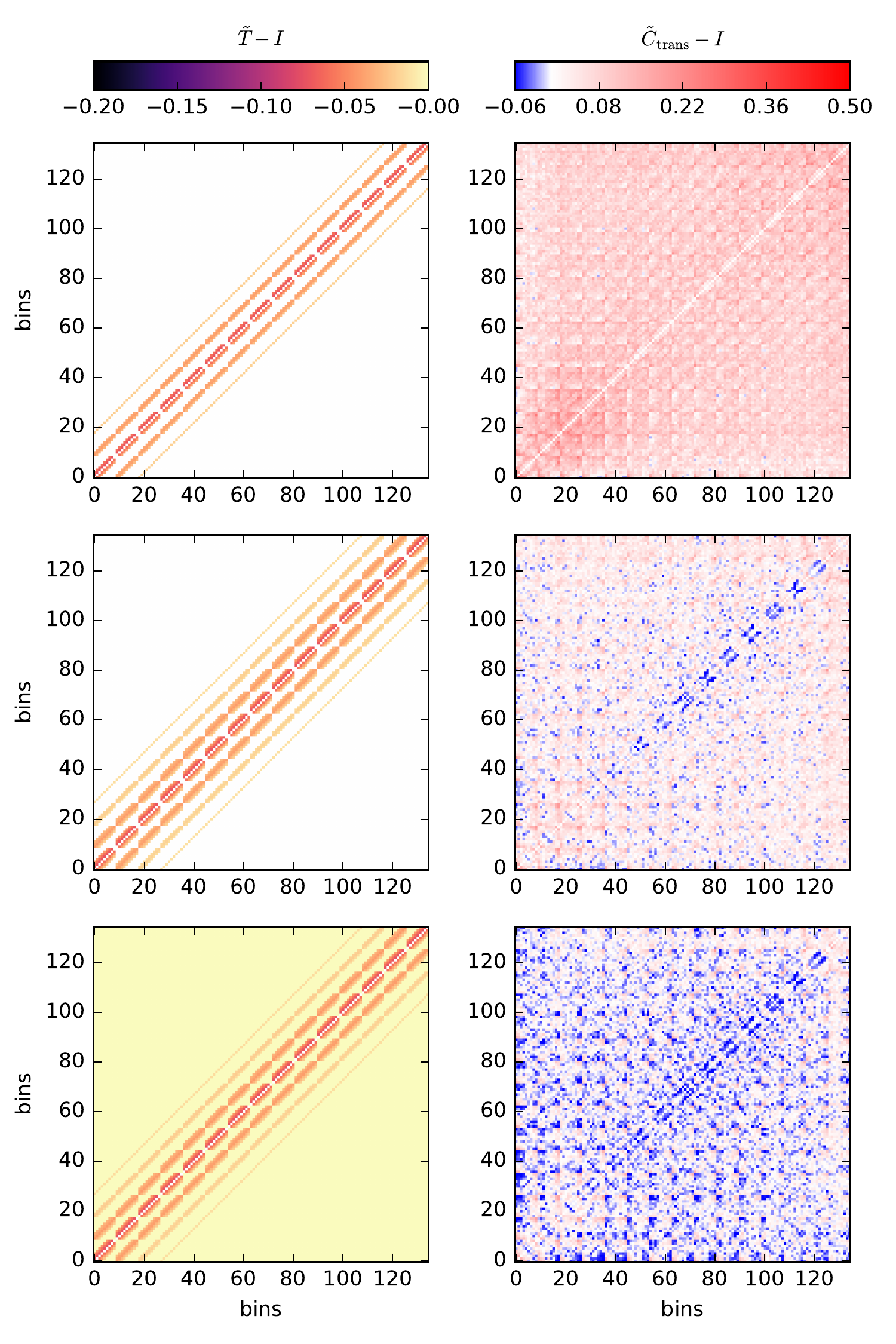}
\vspace{-0.4cm}
\caption{The left panels show 3 different approximations to the Fisher square root matrix of the prewhitened anisotropic 2PCF. The matrix is normalized with the diagonal and then we subtract off the identity matrix to reveal the off-diagonal terms. The top and middle row correspond to Equation~\ref{equ:invsqrt_kernel_2_vals} and Equation~\ref{equ:kernel_3_invsqrt_vals}, respectively. The bottow row adds a small pedestal value to the full transformation matrix. The right panels show the reduced covariance matrix of the transformed anisotropic 2PCF after applying the corresponding transformation matrix. Again, the unity diagonal has been subtracted off to reveal the structure of the off-diagonals.}
\label{fig:cov_invsqrt_xi}
\end{figure*}

We begin by modeling the transformation kernel with the following form 
\begin{equation}
\centering
\begin{array}{*{20}c}
           &       &\times &       &  \\
           &\times &\times &\times &  \\ 
    \times &\times &1      &\times &\times  \\ 
           &\times &\times &\times &  \\ 
           &       &\times &       &  \\
 \end{array}.
 \label{equ:invsqrt_kernel_2}
\end{equation}
Again, the horizontal axis represents the $d_\perp$ axis and the vertical axis represents the $d_\parallel$ axis. Each $\times$ represents a kernel weight to be determined. The corresponding transformation matrix has a central band of width 5 and 2 extra off-diagonal bands on each side. We compute the kernel weights by averaging the Fisher square root matrix along each off-diagonal. We present the weights in Equation~\ref{equ:invsqrt_kernel_2_vals}. 
\begin{equation}
\centering
\begin{array}{*{20}c}
           &       &-0.059 &       &  \\
           &-0.035 &-0.072 &-0.037 &  \\ 
    -0.020 &-0.041 &1      &-0.040 &-0.018  \\ 
           &-0.031 &-0.058 &-0.027 &  \\ 
           &       &-0.038 &       &  \\
 \end{array}.
 \label{equ:invsqrt_kernel_2_vals}
\end{equation}
The resulting transformation matrix is shown in the top left panel of Figure~\ref{fig:cov_invsqrt_xi}. 

The top right panel of Figure~\ref{fig:cov_invsqrt_xi} shows the corresponding transformed covariance matrix. We see dramatic reduction to the off-diagonal terms of the covariance matrix compared to Figure~\ref{fig:cov_xi}. The mean residual on the transformed covariance matrix is 0.089, down ${\sim} 60\%$ from the mean residual of 0.221 in the pre-transform covariance matrix. With this transform, we estimate $\chi^2 = 1415$, a substantial improvement over the pre-transform estimate of 2206.

To combat the residuals, we follow the same modifications as we did for the Cholesky matrix. We first expand the transform kernel to include 3 neighboring bins in all four directions, visualized in the following form 
\begin{equation}
\centering
\begin{array}{*{20}c}
          &       &       &\times &       &       &  \\
          &       &\times &\times &\times &       &  \\ 
          &\times &\times &\times &\times &\times &  \\ 
   \times &\times &\times &1      &\times &\times &\times \\ 
          &\times &\times &\times &\times &\times &  \\ 
          &       &\times &\times &\times &       &  \\ 
          &       &       &\times &       &       &  \\
 \end{array}.
 \label{equ:kernel_3_invsqrt}
\end{equation}
Again, the horizontal axis corresponds to the $d_\perp$ direction, and the vertical axis corresponds to the $d_\parallel$ direction. The weights are againt computed by averaging the corresponding diagonal of the Fisher square root matrix, and we show the resulting values in Equation~\ref{equ:kernel_3_invsqrt_vals}.
\begin{equation}
\centering
\begin{array}{*{20}c}
          &       &       &-0.043 &       &       &  \\
          &       &-0.030 &-0.059 &-0.028 &       &  \\ 
          &-0.017 &-0.035 &-0.072 &-0.037 &-0.015 &  \\ 
   -0.013 &-0.020 &-0.041 &1      &-0.040 &-0.018 &-0.012 \\ 
          &-0.013 &-0.031 &-0.058 &-0.027 &-0.013 &  \\ 
          &       &-0.019 &-0.038 &-0.018 &       &  \\ 
          &       &       &-0.022 &       &       &  \\
 \end{array}.
 \label{equ:kernel_3_invsqrt_vals}
\end{equation}
The middle left panel of Figure~\ref{fig:cov_invsqrt_xi} showcases the modified transformation matrix, and the middle right panel shows the corresponding transformed covariance matrix. We see dramatic reduction to the off-diagonal terms compared to the top right panel, especially in the $d_\perp \sim 2$~Mpc region, with a small ${\sim} 0.1$ residual left. The overall mean residual on the transformed covariance matrix is 0.036, representing a ${\sim} 84\%$ reduction compared to the pre-transform case. The estimated $\chi^2 = 1276$, a further improvement compared to the 2-diagonal model.

Finally, we add a pedestal value to the full transformation matrix and have its value fit to minimize the square sum of the off-diagonal values of the transformed covariance matrix. For our mocks, we get a pedestal value of $-6.90\times 10^{-4}$. The resulting transformation matrix and its corresponding transformed covariance matrix are shown in the bottom row of Figure~\ref{fig:cov_invsqrt_xi}. We see the addition of this pedestal value moderately suppresses the broad residuals in the transformed covariance matrix. The mean residual on the transformed covariance matrix is 0.029, an ${\sim} 87\%$ reduction compared to the pre-transform case and a moderate improvement over not using a pedestal value. Similar to the Cholesky case, the only noticeable residuals are at periodic spots where we are cross-correlating with the edge at high $d_\parallel$, where we suffer from high noise due to small sample size.
With the transformation matrix, we estimate the $\chi^2 = 1326$, which is $7\%$ off from $\chi_{\rm{True}}^2 = 1234$. This also represents a small improvement over the Cholesky case, where we get a $9\%$ deviation from the true $\chi^2$.

However, the $\chi^2 = 1326$ is further off from the true value then the estimated $\chi^2 = 1276$ using the 3-diagonal model without the broad pedestal term. We repeat this test with several different HOD perturbations and find no consistent winner between the two models. Together with the mean residual results, this suggests that the inclusion of a broad pedestal value does not meaningfully improve the model of transformation matrix in the case of the anisotropic 2PCF.

\begin{figure*}
\centering
 \hspace*{-0.4cm}
\includegraphics[width=6in]{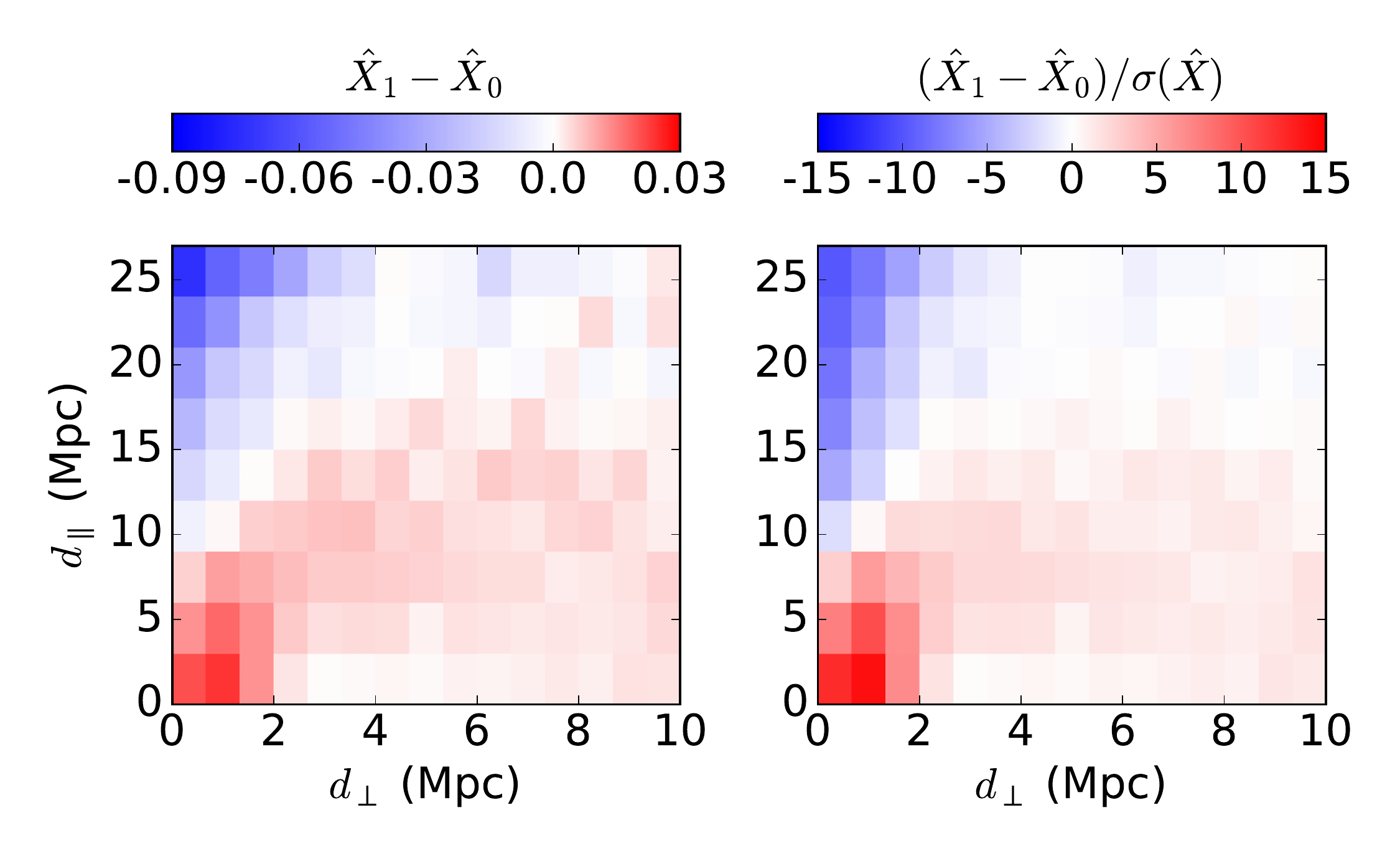}
\vspace{-0.4cm}
\caption{The left panel shows the perturbation to the prewhitened anisotropic 2PCF when we change the satellite distribution parameter $s$ from 0 to 0.2. Again, compared to the perturbations shown in Figure~7 of \citet{2018Yuan}, we see that we have recovered qualitatively the same behavior as the pre-transform anisotropic 2PCF. The right panel shows the corresponding signal-to-noise in each bin, where the noise is computed from the diagonal of the transformed covariance matrix. We can estimate the overall $\chi^2$ by adding up all the bins on the right panel in quadrature.}
\label{fig:xi_smooth_invsqrt}
\end{figure*}

We can now compute the transformed anisotropic 2PCF, $\hat{X}$, using the transformation shown in the bottom left panel of Figure~\ref{fig:cov_invsqrt_xi}. We present the perturbation to $\hat{X}$ when we change the HOD from the baseline HOD to the HOD given by Table~1 of \citet{2018Yuan} and the corresponding signal-to-noise. The results are plotted in Figure~\ref{fig:xi_smooth_invsqrt}. 

The perturbation to the transformed anisotropic 2PCF shown in the left panel of Figure~\ref{fig:xi_smooth_invsqrt} reveals a qualitatively similar pattern to the perturbation to the Cholesky-transformed anisotropic 2PCF shown in in the left panel of Figure~\ref{fig:xi_smooth_cholesky} and that to the pre-transform anisotropic 2PCF shown in Figure~7 of \citet{2018Yuan}. This suggests that the transformation matrix is sufficient compact in scale to preserve the qualitative behaviors of the pre-transform 2PCF. The right panel shows the corresponding signal-to-noise of this perturbation in the transformed 2PCF. The noise $\sigma(\hat{X})$ is computed from the diagonal terms of the transformed covariance matrix.

\begin{table}
\begin{tabular}{ c | c c c}
\hhline {====}
 & $N_{\rm{params}}$ & mean residual & $\chi^2$\\ 
\hline
Cholesky 2 (Eq. \ref{equ:kernel_2_vals})  & 6 & 0.061 & 1253 \\ 

Cholesky 3 (Eq. \ref{equ:kernel_3_vals}) & 12 & 0.029 & 1100\\ 

Cholesky 3+$\epsilon$ (Eq. \ref{equ:kernel_3_vals} + $\epsilon$) & 13 & 0.028 & 1119 \\
\hline 
FS 2 (Eq. \ref{equ:invsqrt_kernel_2_vals}) & 12 & 0.089 & 1415\\ 

FS 3 (Eq. \ref{equ:kernel_3_invsqrt_vals}) & 24 & 0.036 & 1276\\ 

FS 3+$\epsilon$ (Eq. \ref{equ:kernel_3_invsqrt_vals} + $\epsilon$) & 25 & 0.029 & 1326\\
\hline 
\end{tabular} 
\caption{A summary of all the transformations we proposed for the anisotropic 2PCF. The first column lists the names of the transformations and the corresponding equation number. ``FS" stands for Fisher square root, and the number following describes the number of bins we include in each direction. $\epsilon$ signals the use of a pedestal value. The second column and the third column summarizes the number of parameters used to construct the transformation matrix and the resulting mean residual in the transformed covariance matrix. The fourth column lists the estimated $\chi^2$ of the transformed 2PCFs with just the diagonal of the transformed covariance matrix. The 2PCF perturbation is drawn from the HOD perturbation quoted in the first row of Table 1 of \citet{2018Yuan}. The pre-transform covariance matrix has a mean residual of 0.221 for reference. The true $\chi^2_{\rm{True}} = 1234$.}
\label{tab:summary_xi}
\end{table}

Table~\ref{tab:summary_xi} presents a summary of the different transformations we have constructed for the anisotropic 2PCF. The first column shows the names of the transformations and their corresponding equation numbers. The second column shows the number of parameters needed to construct the transformation matrices. The third column shows the mean residual of the transformed covariance matrices. The fourth column shows the $\chi^2$ estimate using just the diagonal of the transformed covariance matrices. The top and bottom three rows show the Cholesky and Fisher square root results, respectively. Comparing to the pre-transform mean residual of 0.221, we see that even the simplest 6 parameter Cholesky-based transformation matrix suppresses the off-diagonal covariances by ${\sim} 72\%$ and returns a substantial better estimate of the $\chi^2$. The inclusion of one more term along each direction in the kernel further reduces the off-diagonal covariances, whereas the introduction of a broad pedestal value does not seem to help much. Comparing the Cholesky cases to the Fisher square root cases, we see that the Cholesky cases and the Fisher square root cases have similar performances while the Cholesky cases introduce about half as many parameters. The ``FS 3+$\epsilon$" case does produce a more accurate $\chi^2$ than the ``Cholesky 3+$\epsilon$ for our example perturbation. However, the two $\chi^2$ only differ by $2\%$ of the true $\chi^2$, and our tests with different HOD perturbations do not reveal any consistent performance difference between the two. We do perfer the Cholesky set of models as they utilize fewer parameters.



\section{Discussion $\&$ Conclusions}
\label{sec:conclusions}

In this paper, we propose methods for decorrelating the projected 2PCF and the anisotropic 2PCF using simple and compact transformation matrices modeled after the Cholesky decomposition and the symmetric square root of the Fisher matrix. For both the projected 2PCF and the anisotropic 2PCF, we have shown the transformed 2PCF still recover the same structure as the pre-transform 2PCF. Thus, the transformed 2PCF can be interpreted in the same way as its pre-transform counterpart. For both the projected 2PCF and the anisotropic 2PCF, we found that the error bars computed from the diagonal of the transformed covariance matrix are much larger than those computed from the diagonal of the pre-transform covariance matrix, confirming that error bars computed from the diagonal of the pre-transform covariance matrix dramatically underestimate the uncertainty in the 2PCF.

For the projected 2PCF, we found that that the Cholesky models consistently outperforms the Fisher square root models in suppressing off-diagonal covariances, and that the addition of a small broad pedestal value helps greatly.
The 4-parameter Cholesky model of the transformation model performs particularly well, suppressing off-diagonal covariances by ${\sim} 95\%$ and returning a $\chi^2$ estimate that is impressively close to the true value. Key test results are presented in Table~\ref{tab:summary_wp}.

For the anisotropic 2PCF, we found that even a simple 6-parameter Cholesky based model can suppress the off-diagonal covariances by ${\sim} 72\%$ and return a good estimate of the $\chi^2$. We found that the Cholesky models and the Fisher square root based models have similar performance but the Cholesky models require fewer parameters. The inclusion of a broad pedestal value does not seem to meaningfully improve the results. Our best Cholesky model including 3 bins in each direction and a pedestal value suppresses the off-diagonal covariances by ${\sim} 87\%$ and yields a $\chi^2$ estimate that is $9\%$ off from the true value. We show the key test results in Table~\ref{tab:summary_xi}.

We propose these transformations as examples where one can perform a simple linear transform to the observed or simulated 2PCF and obtain a decorrelated version where the error bars are more representative of the level of uncertainty. At the same time, these transforms are relatively compact in scale that the transformed statistics still show similar structures and can still be interpreted in the same way as the vanilla 2PCF. The addition of a broad pedestal value does break the compactness in real space, albeit in a simple fashion, resulting in a $10-20\%$ rescaling. We do not have an intuitive physical interpretion of the pedestal value, and hence the model of a pedestal value may need to be revisited in other scenarios.

These transformations also offer an opportunity for data compression in the covariance matrix. Normally with a projected 2PCF in 30 bins, the Fisher matrix has 465 unique elements. However, with Equation~\ref{equ:cholesky_longer_transform}, we have proposed a 4-parameter model for the Fisher matrix of the projected 2PCF. For the anisotropic 2PCF with $15\times 9$ bins, the Fisher matrix would normally have 9180 unique elements. Howevever, Equation~\ref{equ:kernel_3_vals} with an additional pedestal value reduces the number of unique parameters to 13. 
By using these highly regularized models, one would need to generate much fewer mocks to determine the Fisher matrix, reducing the computational cost of these fits. One would then use the parameter fits to run more mocks to obtain the proper error bars. Such a routine not only saves considerable computational resources, but can also avoid the bias from inverting a noisy covariance matrix because we are directly fitting a model for the Fisher matrix instead of the covariance matrix. However, it is possible that our highly regularized model can also lead to bias in the parameter fits, but we defer such discussion to future papers.

To summarize, we propose simple and compact transformation matrices that decorrelate the projected and anisotropic 2PCF and enable accurate presentation of the error bars. These transformation matrices also provide a highly regularized template model for the Fisher matrix, drastically reducing the number of mocks needed to determine the covariance matrix and avoiding the bias due to inverting a noisy covariance matrix. 

\section*{Acknowledgements}
DJE is supported by National Science Foundation grant AST-1313285. DJE is additionally supported by U.S. Department of Energy grant DE-SC0013718 and as a Simons Foundation Investigator.

The \textsc{ABACUS} simulations used in this paper are available at \url{https://lgarrison.github.io/AbacusCosmos}.
 
\bibliographystyle{aasjournal}
\bibliography{biblio}
\end{document}